\documentclass[preprint2,usenatbib]{mnras}
\usepackage{graphicx}
\usepackage {amsmath,amssymb}
\usepackage{color}
\usepackage{aas_macros}
\usepackage{array}

\def\fI{ f_{\rm I}}
\def\fII{ f_{\rm II}}
\def\fIII{ f_{\rm III}}

\begin{document}

\title{The Prevalence of Type III Disc Breaks in HI-rich and Low-spin Galaxies}

\author[Jing Wang et al.]{Jing Wang$^1$ \thanks{Email: hyacinthwj@gmail.com}, Zheng Zheng$^2$, Richard D'Souza$^3$, Houjun Mo$^{4, 5}$, Gyula J\'ozsa$^{6,7,8}$,  \\
\newauthor Cheng Li$^4$, Peter Kamphuis$^{9,10}$, Barbara Catinella$^{11,12}$, Li Shao$^1$, Claudia del P. Lagos$^{11,13}$,\\
\newauthor Min Du$^1$, Zhizheng Pan$^{14}$\\  
$^1$ Kavli Institute for Astronomy and Astrophysics, Peking University, Beijing 100871, China\\
$^2$ National Astronomical Observatories, Chinese Academy of Sciences, A20 Datun Road, Chaoyang District, Beijing 100012\\
$^3$ Department of Astronomy, University of Michigan, 1085 South University Avenue, Ann Arbor, MI 48109-1107, USA\\
$^4$ Tsinghua Center for Astrophysics and Department of Physics, Tsinghua University, Beijing 100084, China\\
$^5$ Department of Astronomy, University of Massachusetts Amherst, MA 01003, USA\\
$^6$ Rhodes University, Department of Physics and Electronics, Rhodes Centre for Radio Astronomy Techniques \& Technologies, \\
 PO Box 94, Grahamstown, 6140, South Africa\\
$^7$ Argelander-Institut f\"ur Astronomie, Auf dem H\"ugel 71, D-53121 Bonn, Germany\\
$^8$ SKA South Africa Radio Astronomy Research Group, 3rd Floor, The Park, Park Road, Pinelands, 7405, South Africa \\
$^9$ National Centre for Radio Astrophysics, TIFR, Ganeshkhind, Pune 411007, India\\
$^{10}$ Astronomisches Institut der Ruhr-Universit\"at Bochum (AIRUB), Universit\"atsstrasse 150, 44801 Bochum, Germany \\
$^{11}$ International Centre for Radio Astronomy Research, M468, The University of Western Australia, Crawley, WA 6009, Australia\\
$^{12}$ ARC Centre of Excellence for All Sky Astrophysics in 3 Dimensions (ASTRO 3D)\\
$^{13}$ Australian Research Council Centre of Excellence for All-sky Astrophysics (CAASTRO), 44 Rosehill Street Redfern, NSW 2016, Australia.\\
$^{14}$ Purple Mountain Observatory, Chinese Academy of Sciences, 2 West-Beijing Road, Nanjing 210008, China\\
}
\maketitle

\begin{abstract}
We investigate the origin of the type III (up-bending) discs based on a sample of 1808 galaxies from SDSS and a sub-sample of 286 galaxies with H{\textsc I} data from ALFALFA. 
We examine how the type III fraction $\fIII$, the fraction of disc galaxies which host up-bending disc breaks, depends on other galactic properties. We confirm that $\fIII$ strongly depends on the stellar concentration of galaxies. We find that H{\textsc I}-rich galaxies with low spins tend to have significantly more type III disc breaks than control galaxies, which are matched in concentration and stellar mass. This trend is independent of the existence of strong bars or environment of the galaxies. This result is broadly consistent with predictions from theoretical simulations, and indicates in-situ star formation fueled by gas accretion to be an important mechanism that builds the outer discs of type III galaxies.

\end{abstract}


\section{Introduction}
\label{sec:introduction}
The radial surface brightness profiles of galaxy discs do not always follow single exponential functions, but commonly break into piece-wise exponential functions with different scale-lengths \citep{Pohlen06, Erwin08, Gutierrez11}. For late type galaxies, more than half of the discs have their scale-lengths transiting to a smaller one in the outer regions (down-bending, or type II discs). Many of the remaining discs ($\sim30\%$ of all discs) go the other way around (up-bending, or type III discs) and only a small fraction ($\sim$10\%) of discs show one single scale-length within the whole measurable radius range (type I discs) \citep{Pohlen06}. There are three major processes that add stars to the outskirts of galactic discs: stellar migration from the inner discs \citep{Sellwood02}, accretion of stars stripped from satellite galaxies \citep{Cooper13}, and in-situ star formation from the accretion of gas \citep{Mo98}. Their relative importance is supposed to differ between type II and III galaxies \citep[see][for a review]{Debattista17}.

Although a small fraction of the type II breaks may be caused by orbital resonance related to bars \citep{Erwin08, Minchev12, MunozMateos13}, migration induced by spiral arms or bars may dominate the origin of stars in the outer discs for most of the type II galaxies. One strong evidence is the U-shaped average age profiles commonly found in type II galaxies, with an age up-turn close to the break radius \citep{deJong07, Bakos08, RadburnSmith12, Zheng15, Dale16}. Beyond the break radius, young stellar populations show steeper surface density profiles \citep{MunozMateos09, RadburnSmith12}, but they share the break radii with the old stellar populations \citep{deJong07,MartinezLombilla18}. These results for type II galaxies can be coherently understood in a picture where the break is caused by the suppressed star formation efficiency in low surface density regions, possibly related to H{\textsc I} warps \citep{SanchezBlazquez09}; the declining parts of the U-shape age profiles reflect the inside-out growth of discs via star formation; and the rising parts beyond the break radius are dominated by stars originating from the inner discs, with old stars having migrated further than young stars \citep{Roskar08}. Other mechanisms can also contribute to the stars beyond the break radius, but may either have a lower significance or tend to transform the type II disc into other types \citep{RuizLara17}. 

The origin of stars in the type III outer discs is more complex. Theoretically, stars originating from small and circular orbits can be efficiently scattered to large and elongated orbits by bars \citep{Brunetti11}; this kind of migration can turn a type I or II galaxy into a type III galaxy \citep{RuizLara17, Herpich17}, and the effect might be strengthened in low-spin halos \citep{Herpich15}. Observationally, type III galaxies indeed tend to have shorter scale-lengths, and potentially lower spins, than the type II galaxies \citep{Pohlen06, Gutierrez11}. On the other hand, the direct link between type III disc breaks and bars has not been systematically studied, and the effect of bars may not explain the formation of type III discs in galaxies without bars.
In numerical simulations, mergers may help form a type III break via three mechanisms \citep{Younger07, Borlaff14}. Firstly, the inner stars of galaxies may gain angular momentum from the torques of tidal forces and move outward\citep{Younger07}, secondly the gas inflow driven by galactic interactions could steepen the inner stellar surface density profile increasing the contrast with the outer discs \citep{Younger07}, and thirdly the stars stripped from the satellite galaxies during the mergers also add to the outer discs \citep{Cooper13}. The average colour profile observed for type III galaxies flattens in the outer regions \citep{Bakos08, Zheng15}, which is consistent with the accretion of a significant fraction of satellite stars \citep{RuizLara16}. On the other hand, the effect of mergers should strongly depend on the environment of galaxies, but a correlation between disc breaks and the environment is not observationally confirmed \citep{Laine14, Maltby12}. 
 In-situ star formation may also play a role in the formation of type III discs. In galaxy formation models with a $\Lambda$CDM cosmology, gas accretion leads to the inside-out formation of discs \citep{Mo98}. It may be natural to expect the in-situ star formation to be strong enough to produce an up-bending type III disc when the gas accretion is strong and the galaxies are highly H{\textsc I}-rich \citep{Minchev12}. Observationally, H{\textsc I}-rich galaxies tend to have more star-forming outer discs beyond the half-light radius than other galaxies \citep{Wang11}, but it is unclear whether this trend extends beyond the break radius. 
 Hence the stars in the outer discs of type III galaxies may be accreted, migrated and directly formed; the formation theory of type III breaks is much less conclusive than that of type II discs.

Understanding the role of each process in shaping the optical outer discs, provides important constraints on the formation models of disc galaxies. This is because, although the stars in the outskirts represent only a small fraction of the total stellar mass in galaxies, the related processes of stellar accretion and gas accretion are all major contributors to the assembly of the total stellar mass. Stellar migration does not change the total stellar mass of galaxies, but it changes their age and metallicity profiles. Type II and III galaxies do not just differ in their outer discs. Type III galaxies have more compact inner discs \citep{Pohlen06}, and they are more common among early-type galaxies than type II discs \citep{Gutierrez11}. The scale-lengths of the inner discs of type II galaxies decrease with observing wavelengths from ultraviolet to optical, while for type III galaxies the trend reverses \citep{Herrmann13}. Because the breaks typically start close to or beyond the optical R$_{90}$ (radius that enclose 90\% of the total light) \citep{Zheng15}, these results imply that the different ways of mass assembly in the outer discs are likely coupled to the different ways of forming the inner discs and maybe the whole galaxy. Although simulations are reaching resolutions and the required sophistication to simulate realistic disk galaxies, they have not yet explored the nature of different disks exhaustively, so it still is an open question whether current simulations reproduce the correct frequency of type I/II/III disks as well as their correlation with galaxy properties.

Most of the observations of type III breaks have been limited to the optical to mid-infrared bands, while new insights may be obtained from observations of the H{\textsc I} gas. The spectrum of H{\textsc I} provides at least two new parameters that are key to gain insights in the puzzle of different formation theories. The H{\textsc I} mass (or H{\textsc I}richness) is directly related to gas accretion. Because the H{\textsc I} discs typically extend further than the optical discs in star-forming galaxies \citep{Swaters02}, the line width of the H{\textsc I} spectrum (corrected for inclination) provides a good estimate of the maximum rotational velocity and thereby part of the spin parameter \citep{Huang12}, which may be related to the stellar migration effects driven by bars \citep{Herpich15}. In this paper, we make use of the H{\textsc I} data from ALFALFA \citep[Arecibo Legacy Fast ALFA]{Giovanelli05}, and search for clues of possible mechanisms that contribute to the formation of type III breaks. The paper is organized as follows.  Section~\ref{sec:data} introduces the data and parameters to be studied, and describes the procedure for identifying strong bars and type II and III breaks. Section~\ref{sec:result} investigates how type III breaks are related to the H{\textsc I}-richness, the spin parameter, the existence of bars and the galactic environment. Section~\ref{sec:discussion} discusses the interpretation of our results, and finally, section~\ref{sec:conclusion} concludes. 
Throughout this paper, we assume a Chabrier IMF \citep[initial mass function,][]{Chabrier03}, and a $\Lambda$CDM cosmology with $\Omega_{m}=0.3$, $\Omega_{lambda}=0.7$ and $h=0.7$. 

\section{Data and analysis}
\label{sec:data}
\subsection{The optical data}
\label{sec:SDSSdata}
\subsubsection{SDSS and the starting sample}
The Sloan Digital Sky Survey \citep[SDSS][]{Abazajian09} comprises a wide-field image survey in the $u$, $g$, $r$, $i$ and $z$ bands, and an optical spectroscopic follow-up survey. The images have a pixel size of 0.396$\arcsec$ and typical point spreading function (PSF) full-width half-maximum (FWHM) width of 1.3$\arcsec$. The spectra cover the wavelength range of 3800-9200 $\AA$. We use the images from Data Release 10 \citep[DR10,][]{Ahn14}.

We use the MPA$\slash$JHU catalog \footnote{http://wwwmpa.mpa-garching.mpg.de/SDSS/DR7/} of galaxies detected in DR7 of SDSS. This catalog includes photometric measurements of magnitudes and sizes, and provides estimates of redshifts and stellar masses ($M_*$) for over 900,000 galaxies. Based on the MPA$\slash$JHU catalog, we further calculate the following parameters that are widely used to indicate galaxy properties \citep[e.g.]{Kauffmann03,Wang11}:
\begin{itemize}
\item $R_{90}/R_{50}$, the ratio between the 90-percent and half-light circular radius, also termed concentration, measured in the $r$-band.
\item $b/a$, the ratio between the minor and major axis of the 25 mag arcsec$^{-2}$ isophote, measured in the $g$-band and related to the inclination of galaxies.
\item $\mu_*$, the average stellar mass surface density within the effective radius of the SDSS $z$-band.
\item $\Delta(g-i)$, the difference of the $g-i$ colours between the outer and inner discs of the galaxies, where the inner discs are within $R_{50}$ and the outer discs are between $R_{50}$ and $R_{90}$.
\end{itemize}
Galaxies with high values of $R_{90}/R_{50}$ or $\mu_*$ tend to concentrate a large fraction of stellar mass in their central bulges. Face-on galaxies have $b/a$ close to unity. More negative values of $\Delta(g-i)$ indicate more active star formation on the outer regions of galaxies.

We select all the galaxies which have a redshift between 0.01 and 0.03, a central concentration index $R_{90}/R_{50}$ between 1.8 and 2.8, $b/a$ above 0.5 and $M_*$ between 10$^{10}$ and 10$^{11.5}~M_{\odot}$. According to the properties indicated by the parameters entering the selection criteria \citep{Kauffmann03, Nair10}, we have selected a sample of face-on and disc-like massive galaxies at low redshift, in total 2283 galaxies. If we more conservatively select disc galaxies by requiring $R_{90}/R_{50}<2.6$ the major results (correlations) presented in this paper do not change. We also confirm in Section~\ref{sec:bulgemask} by analyzing the radial distribution of light that the majority of these galaxies are disc dominated. We inspect the SDSS images of all these galaxies and exclude 320 galaxies with bright neighbor sources that significantly contaminate the light in the outer discs of these galaxies. The remaining 1963 galaxies comprise the {\it starting sample} of this paper.

\subsubsection{Deriving surface brightness radial profiles and the main sample}
\label{sec:getSBprof}
We use a slightly modified version of the photometric pipeline outlined in \citet{Wang12} to obtain radial surface brightness (SB) profiles in the SDSS $r$-band for galaxies in the {\it starting sample}. We use the segmentation images produced by SExtractor \citep{Bertin96} to make mask images for the $r$-band images. We make two mask images for each $r$-band image. The first mask image has all galaxies masked. The second mask image has all except for the target galaxy masked. We estimate the sky background in two steps. In the first step, we estimate an initial background value from the original image by calculating the 3-$\sigma$ clipped mean value of the pixels that are not masked by the first mask image. In the second step we remove the residual background value when deriving the SB profile. We use the initial background value and the second mask image as input for the IRAF task ELLIPSE \citep{Jedrzejewski87}, and run the task twice to obtain the radial SB profiles from the $r$-band images. 

 In the first run of ELLIPSE, we allow the position angles and ellipticities (1$-b/a$) to vary as a function of radius. We take the outmost reliably fitted ellipse as the intrinsic (global) projected shape of the disc. In the second run, we fix the position angle and ellipticity to the outmost values determined in the first run. By doing so, we are able to reach lower azimuthally averaged SBs than what was obtained in the first run. We measure all profiles out to 4 times $R_{25}$ (major axis of the 25 mag arcsec$^{-2}$ isophote in the $g$-band). The profiles typically become flattened beyond $2R_{25}$, indicating that they reach the SB of the residual background.
 
 We then estimate the residual background value from the SB profile.
 We calculate the mean and standard deviation ($\sigma_{back}$) of the flattened outer profile beyond $2R_{25}$. We take the mean value as the residual background value and subtract it from the SB profile. We take 6$\sigma_{back}$ as the SB depth that can be reached by the image, and use it to determine the outmost radius of the SB profile ($r_{out}$). So that a fluctuation of 1$\sigma_{back}$ corresponds to a change of $\sim$0.17 mag arcsec$^{-2}$ in the SB value. An example of the estimate of the residual background values can be found in Figure~\ref{fig:example_backsub} in Appendix.

\begin{figure*} 
\includegraphics[width=20cm]{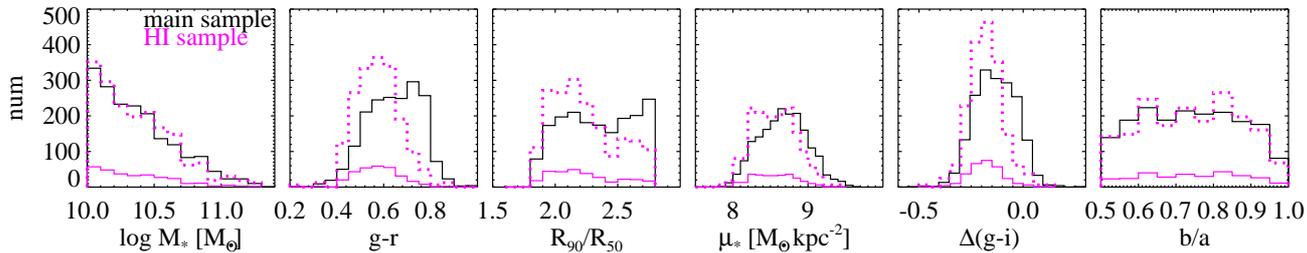}
\caption{The distribution of $M_*$, $g-r$, $R_{90}/R_{50}$, $\mu_*$, $\Delta(g-i)$ and $b/a$ for galaxies. Black histograms are for the main sample and magenta colours are for the H{\textsc I} sample. The magenta histograms in solid lines are directly calculated from the H{\textsc I} samples, and the magenta histograms in dotted lines are the magenta solid histograms scaled up by the ratio of the sample sizes between the main sample and the H{\textsc I} sample. }
\label{fig:galprop}
\end{figure*}

ELLIPSE failed for $\sim7.9\%$ of the galaxies, and finally we are able to reliably measure SB profiles for 1808 galaxies, which make the {\it main sample} of our study. Figure~\ref{fig:galprop} shows the distributions of optical properties for this sample. 

{\rm $r_{out}$ of our sample ranges from 15.44 to 63.76 arcsec (5 and 95 percentiles), with a median value of 31.28 arcsec. Physically, $r_{out}$ ranges from 8.2 to 29.4 kpc (1.62 to 2.87 $R_{90}$), with a median value of 16.1 kpc (2.1 $R_{90}$). Based on the previous studies which derived the radial ellipticity profiles from the stacked images of hundreds and thousands of galaxies \citep{Dsouza14, Zheng15}, the averaged ellipticity remains high at the radius within $r_{out}$ of galaxies from the main sample. Hence the radial range studied in this paper is not significantly contaminated by light from the halo stars.

The SB depth (6$\sigma_{back}$) ranges from 25.49 to 26.56 (5 and 95 percentiles) and has a median value of 26.06 mag arcsec$^{-2}$. Hence we are able to detect the typical surface brightnesses (23 and 25 mag arcsec$^{-2}$ respectively) at the break radius of type II and III galaxies \citep{Pohlen06}.  }

\subsubsection{Masking the bulge dominated regions}
\label{sec:bulgemask}
 We identify the bulge dominated region of each profile from the second-order derivative of the SB profile ($\frac{{\rm d^2}\Sigma}{{\rm d^2}r}$). The method is based on the different characteristic profile shapes between discs and bulges \citep{Kormendy16}. If we consider a galaxy with a type I disc and a central bulge, the SB profile of the disc dominated region can be described by an exponential function of the radius, while the SB profile of the bulge dominated region rises much more steeply towards the center than any exponential functions. In the logarithm space, the first-order derivative of the SB profile  ($\frac{{\rm d}\Sigma}{{\rm d}r}$) remains almost unchanged in the disc dominated region, while varies continuously as a function of radius in the bulge dominated region. The second-order derivative $\frac{{\rm d^2}\Sigma}{{\rm d^2}r}$ will be close to zero in the disc dominated region, while deviating significantly from zero in the bulge dominated region. Type II and III galaxies may have non-zero values of $\frac{{\rm d^2}\Sigma}{{\rm d^2}r}$ near the break radius, which however do not significantly affect the method.

In practice, we smooth the SB profile with a boxcar kernel of 5 pixels ($\sim$2 arcsec) before deriving the derivative profiles. After obtaining the second-order derivative profile, we calculate the 3$\sigma$-clipped standard deviation of $\frac{{\rm d^2}\Sigma}{{\rm d^2}r}$ ($\sigma_{deriv}$). We identify the beginning of the disc region ($r_{in}$) as the smallest radius where the absolute value of $\frac{{\rm d^2}\Sigma}{{\rm d^2}r}$ drops below $\sigma_{deriv}$. An example of this procedure for masking the bulge dominated regions can be found in Figure~\ref{fig:example_bulge} in Appendix. 

We only work on the radial regions between $r_{in}$ and $r_{out}$ hereafter. $r_{in}/r_{out}$ have a range between 0.08 and 0.31 (5 and 95 percentiles of the distribution) and a median value of 0.17 in the main sample, which supports the selection of the sample to be mainly disc dominated galaxies.

\subsubsection{Identification of disc break types}
\label{sec:breakclassify}
We use the model from \citet[][E08 hereafter]{Erwin08} to describe the SB profile of a disc which breaks into an inner and an outer part:
\begin{equation}
I(r) = S~I_0  e^{\frac{-r}{\gamma}}[1+e^{\alpha(r-r_b)}]^{\frac{1}{\alpha}(\frac{1}{\gamma}-\frac{1}{\beta})}
\end{equation}
$I_0$ is the central SB of the inner exponential disc, $S$ is a scaling constant to force $I(0)=I_0$, $r_b$ is the break radius, $\gamma$ is the scale-length of the inner exponential disc ($r_{s,in}$) and $\beta$ is the scale-length of the outer exponential disc ($r_{s,out}$). $\alpha$ describes the smoothness of the inner exponential profile transiting into the outer exponential profile. It has a range of 0.1 to 1, with larger values indicating sharper transitions. 

Below we describe our procedure for identifying disc breaks.
 As part of this procedure, we fit an E08 model to the SB profile of the disc region (between $r_{in}$ and $r_{out}$) of a galaxy.  The disc is classified type I if 
\begin{enumerate}
\item $1/1.1<r_{s,in}/r_{s,out}<1.1$, 
\item the outer disc extends for at least 5.2 arcsec (4 times the FWHM of the PSF, equivalent to the total length of four independent radial elements); in other words, $r_b$ is at least 5.2 arcsec away from $r_{out}$.
\end{enumerate}

Otherwise the disc is classified broken, the procedure takes the best-fit E08 model as the ``original model'', and starts an iterative process to search for more break radii beyond the break radius $r_b$:
\begin{enumerate}
\item The procedure sets the break radius $r_b$ from the original model to be a new starting point, $r_{in}'$, and fits the E08 model to the profile between $r_{in}'$ and $r_{out}$. The new best-fit model has a new set of parameters including $r_b'$, $r_{s,in}'$ and $r_{s,out}'$.
\item If $1/1.1<r_{s,in}'/r_{s,out}'<1.1$, the procedure stops with no new breaks identified, otherwise it goes to the next step to assess the necessity of accepting this new model.
\item Because the new model introduces additional parameters, the procedure performs the F-test to compare the new model with the original model. When the P value for the F-test of the overall significance is less than 0.0027, the procedure rejects the null-hypothesis that the two models are equal. In other words, when there is more than 99.73\% ($\sim3\sigma$) confidence, the procedure concludes that the new model is a better fit to the data than the original model. If the new model is accepted, the procedure goes to the next step, otherwise it stops with no new breaks identified. 
\item The break radius $r_b$ from the new model needs to be more than 5.2 arcsec away from $r_{out}$. If so, the new model is accepted as a new original model, its break radius is reliably identified, and the procedure goes back to step (i).  Otherwise, the procedure stops with no new breaks identified. 
\end{enumerate}

This iterative procedure of searching for disc breaks is motivated by the observational fact that many galaxy discs have more than one break (E08). 28.7\% and only 2.9\% of the galaxies from our main sample are identified to have two and more than two disc breaks respectively. In the following part of this paper, we only study the properties of the outmost breaks, and all the break-related parameters including $r_b$, $r_{s,in}$ and $r_{s,out}$ are for the outmost breaks. Galaxies which have $r_{s,out}/r_{s,in}<1.1$ and $r_{s,out}/r_{s,in}>1.1$ are classified as type II and III respectively.  

In Figure~\ref{fig:example_prof}, we show examples of type I-III galaxies each of which has one disc break. We also show two examples of type III galaxies which have more than one break. 
 
 We identify 229, 754 and 825 type I-III galaxies, respectively, in the main sample, corresponding to type I-III fractions (referred to as $\fI$, $\fII$ and $\fIII$ respectively hereafter) of 12.7$\%$, 41.7$\%$ and 45.6$\%$. We show in Section~\ref{sec:appendix_breakprop} that the average properties of the disc breaks are consistent with those found in the literature.
 
 We notice that our type II and III fractions are lower and higher, respectively, than some previous studies. The difference is likely due to the different selection effects of the samples, or different procedures for identifying breaks.  For example, \citet{Pohlen06} selected a sample of nearby (D$<$46 Mpc) galaxies with the criteria $b/a>0.5$,$M_B<-18.4$ mag \citep[corresponding to a $M_*>10^{9.13}~M_{\odot}$, assuming a $B-V=0.4$ and the mass-to-light ratio from][]{Bell03}, and Hubble type $2.99<T<8.49$ \citep[roughly corresponding to a $R_{90}/R_{50}<2.5$,][]{Nair10}. They reported type I-III fractions of 10\%, 60\% and 30\% for nearby spiral galaxies. The difference in measured fractions is likely caused by different sample ranges in $R_{90}/R_{50}$ and $M_*$. If we select galaxies with $R_{90}/R_{50}<2.5$ from our main sample, the type I-III fractions change to 10.2\%, 53.6\% and 36.2\%, much closer to the fractions of \citet{Pohlen06} than the main sample. To give another example, \citet[][Z15 hereafter]{Zheng15} selected SDSS galaxies with $b/a>0.5$, the fraction of $r$-band  light contributed from a de Vaucouleurs component fracDev\_r$<0.7$, and the $r$-band petrosain radius petroRad\_r$>5$ arcsec. They reported a type I-III fraction of $\sim$18\%, 59\% and 23\%. The Z15 sample have a similar $R_{90}/R_{50}$ (1.8 to 2.8), but much larger $M_*$ range ($10^{8.5}$-$10^{11.5}~M_{\odot}$) than the main sample of this paper. Z15 also used a different radius range (0.3-2 $R_{90}$) to search for the breaks. If we use the same searching radius range as Z15, the type I-III fractions change to 13.3\%, 51.0\% and 35.8\%, much closer to the results of Z15.

\subsubsection{Identification of bars} 
 We identify strong bar structures through analyzing the position angle and ellipticity profiles produced in the first run of the ELLIPSE procedure. If a bar exists, the radial ellipticity profile rises within the radius of the bar, reaches its maximum at the end of the bar, and is followed by an abrupt drop beyond the bar; in the meantime, the position angle profile does not vary much within the radius of the bar. We refer to \citet{Wang12} for more details of the bar identification procedure. \citet{Wang12} also showed that this method can reliably identify strong bars (ellipticity$>0.5$, length$>$2 kpc) from the SDSS images for galaxies with $M_*>10^{10}~M_{\odot}$ and $z<0.03$. 23.8\% of the galaxies in the main sample are identified to host strong bars. This fraction ($f_{bar}$) is consistent with what was reported in \citet{Wang12}. 10.8\%, 45.9\% and 43.3\% of the galaxies with strong bars (barred galaxies hereafter) host type I-III disc breaks respectively.

\subsubsection{The group catalog}
We use the SDSS group catalog published by \citet{Lim17} to study the environmental dependence of type III galaxies. The halo-based group finder was improved upon the previous versions presented in \citet{Yang05} and \citet{Yang07}.  The reliability of the finder is above 90\% as tested against mock catalogs from cosmological simulations \citep[see][for details]{Lim17}. We use the SDSS group catalog which is constructed with galaxies which have spectroscopic redshifts, and uses stellar mass as the proxy to estimate halo masses \footnote{We also test with the other group catalogs from \citet{Lim17} which are constructed with all SDSS galaxies, or use luminosity as the proxy to estimate halo masses, but the trends presented in this paper do not significantly change.}.

We cross-match our samples with the group catalog, requiring the matched galaxies to have a projected distance within 3 arcsec (SDSS has a spatial resolution of 1.4 arcsec) and a redshift difference within 0.001. This results in a sample of 1228 galaxies (referred to as the {\it environment sample} hereafter), including 733 central galaxies and 495 satellite galaxies. 103 galaxies are in massive groups with a halo mass $M_{\rm halo}>10^{12.5}~M_{\odot}$. The environmental properties used in this paper include the identification of central and satellite galaxies, the $M_{\rm halo}$ and the distances of satellites to the central galaxies. We expect central galaxies and galaxies in large groups to have experienced a richer merger history and hence accreted more stars than satellite galaxies and galaxies in small groups; we also expect that in relatively small groups, satellite galaxies at a small distance to the group center to be affected more by galactic interactions than satellite galaxies at a large distance \citep{Blanton09}.

\subsection{The HI and UV data}
\label{sec:HIdata}
The Arecibo Legacy Fast ALFA \footnote{http://egg.astro.cornell.edu/index.php/} \citep[ALFALFA]{Giovanelli05} is a blind H{\textsc I} survey performed at the single-dish radio telescope Arecibo. It provides H{\textsc I} masses and velocity widths for detected galaxies in the surveyed sky region of 7000 deg$^2$.  The Galaxy Evolution Explorer \citep[GALEX,]{Martin05} is an ultraviolet space telescope that have obtained images in the NUV (near ultraviolet, at $\sim$2100$\AA$) and FUV (far ultraviolet, at $\sim$1510$\AA$) bands for nearly the whole sky except for the sky regions close to the Milky Way. We obtain the photometric measurements of these images from the Mikulski Archive for Space Telescopes (MAST). 

We match our SDSS main sample with the ALFALFA70 catalog (70\% data release of ALFALFA) and GALEX DR8 (Data Release 8). There are 294 galaxies covered by all the three databases. We call them the {\it H{\textsc I} sample}. From Figure~\ref{fig:galprop} we can see that comparing to the main sample, the H{\textsc I} sample is biased toward the galaxies with blue $g-r$ colours, low $R_{90}/R_{50}$, low $\mu_*$ and negative $\Delta(g-i)$. These differences are consistent previous studies \citep{Catinella10, Huang12}. Hence with the H{\textsc I} sample we are investigating star-forming disc galaxies. The type I-III fractions are 9.5\%, 56.1\% and 34.3\% respectively. 

We calculate the following key parameters for the H{\textsc I} sample:
\begin{itemize}
\item $f_{\rm HI}$, the H{\textsc I} mass fraction, calculated as  $f_{\rm HI}=\log M_{\rm HI}/M_*$. 
\item $\Delta f_{\rm HI}$, the H{\textsc I} excess, calculated as the difference between the observed f$_{\rm HI}$ and the averaged f$_{\rm HI}$ of galaxies which have the same NUV$-r$ and $\mu_*$. 
\item $\lambda$, an observational proxy for the spin parameter of the dark matter halo, calculated as 
\begin{equation}
\lambda=21.8\frac{r_{s}[kpc]}{(V_{rot} [km~s^{-1}])^{3/2}}, 
\end{equation}
following Equation 9 of \citet{Hernandez07}. 
We use $r_{s,in}$ for $r_s$ for type II and III galaxies, and the global scale-length for type I galaxies. $V_{rot}$ is estimated as $W_{50}/2/\sin{i}$, where $W_{50}$ is the full-width half-maximum line-width of the H{\textsc I} spectrum. i is the inclination of the optical discs estimated from the axis ratio in the $r$-band, with the equation  $\cos{i}=\sqrt{ ( (b/a)^2-q_0^2)/(1-q_0^2) }$\citep{Holmberg58}, where $q_0$ is the disc thickness and we assume $q_0\sim0.18$ \citep{Courteau97}.
\end{itemize}

We note that \citet{Hernandez07} only applied the equation for calculating $\lambda$ to galaxies with an $r$-band magnitude $-22.5<M_r<-20$. They use the Tully-Fisher relation from Barton et al. (2001) to derive $V_{rot}$, and thus, the calculation is only valid in the magnitude range above. However, because here we estimate $V_{rot}$ directly from the H{\textsc I} line width, we are not limited to this magnitude range. Nevertheless, 95.6\% of the galaxies from the H{\textsc I} sample are in the range $-22.5<M_r<-20$.

We also note that by using $r_{s,in}$ for $r_s$ for type II and III galaxies, we have assumed that the inner disc (within $r_b$) is first formed conserving the spin of the dark matter halo (as predicted by the cosmological models of galaxy formation of \citet{Mo98}, and confirmed to describe the average behavior of galaxies in hydrodynamical simulations by \citet{Lagos17}), and the outer disc (beyond $r_b$) is later assembled via other processes; although the inner scale-lengths may have evolved with the radial movement of gas and stars, on average they still carry information about the initial conditions of their formation. These assumptions are supported by previous studies which found good consistency between the theory and the observation when relating $R_{50}$ to the dark matter spin of the galaxies \citep{Kauffmann03, Shen03}. Nevertheless, we warn the reader that the $\lambda$ parameter calculated in this paper should be viewed only as a proxy for the real spin parameter.

A systematic uncertainty in the calculation of $\lambda$ comes from the assumed fixed intrinsic thickness $q_0$ of 0.18. The thickness of disc galaxies may vary between 0.1 and 0.2 as a function of the Hubble type \citep[e.g.][]{Haynes84}. However because we have selected galaxies with relatively large $b/a$ ($>$0.5), changing $q_0$ from 0.18 to 0.1 or 0.2 only result in a small change ($<1\%$) in the values of $\cos{i}$ and $\sin{i}$.

Galaxies which have $\Delta f_{\rm HI}>0$ tend to have more gas than what is needed to maintain their star forming status with respect to other galaxies, so they are referred to as the H{\textsc I}-rich galaxies. Galaxy discs formed in high-$\lambda$ halos are less compact than other discs \citep{Mo98}. The H{\textsc I} sample has a median $\lambda \sim0.032$, and we use this value to divide the sample into low- and high-$\lambda$ sub-samples.

\section{Results}
\label{sec:result}

\subsection{The dependencies of $\fI$, $\fII$ and $\fIII$ on optical, global galaxy properties}
\label{sec:optprop}
 Z15 showed that  $r_b$ is close to or beyond $R_{90}$ for most of the type II and III galaxies (confirmed in Figure~\ref{fig:breakprop} in the Appendix). The break radius divide the discs into an inner region that contain the bulk mass of the galaxy and the faint outskirts. This section thus investigates the relation between disc break types and some other optical properties measured for the relatively inner region or the whole galaxy. The main goal is to identify the parameters that have the strongest correlations with disc break types, so that in the following sections we can ``control'' these parameters when investigating the other possible mechanisms that affect the formation of type III breaks. 

\begin{figure*} 
\includegraphics[width=14cm]{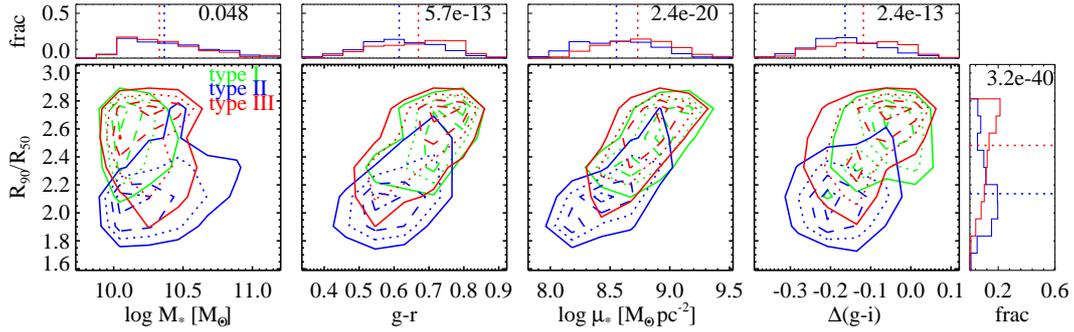}
\caption{The distribution of type I, II and III galaxies in the parameter space of $R_{90}/R_{50}$, $M_*$, $g-r$, $\mu_*$, and $\Delta(g-i)$. The green, blue and red colours are for type I, II and III type galaxies respectively. The solid, dotted, dashed and dash-dotted curves show the 90, 70, 50 and 30\% of galaxy distributions. The dashed lines show the median value of the histograms. In each panel that shows histograms, we denote in the corner the K-S (Kolmogorov-Smirnov) test probability \citep{Press92} which indicates the similarity between the type II and III sub-samples in the property distribution. }
\label{fig:property2d}
\end{figure*}

Figure~\ref{fig:property2d} shows how galaxies with different break types are distributed in the parameter space of $R_{90}/R_{50}$, $M_*$, $\mu_*$, $g-r$ and $\Delta(g-i)$. Type I and III galaxies typically occupy similar regions in these plots. They have on average slightly lower M$_*$, higher $\mu_*$, redder $g-r$, and less negative $\Delta(g-i)$, but the most significant difference from the type II galaxies is evident in the distribution of $R_{90}/R_{50}$. 

Though type II and III galaxies differ in almost all the properties investigated above, Z15 found that the type fractions are much more strongly correlated with $R_{90}/R_{50}$ than with the other parameters. Because our procedure for identifying breaks is different from that in Z15 (see \ref{sec:breakclassify}), we check the trend found by Z15 with our data, and search for possible secondary dependences when fixing the $R_{90}/R_{50}$.

\begin{figure*} 
\includegraphics[width=14cm]{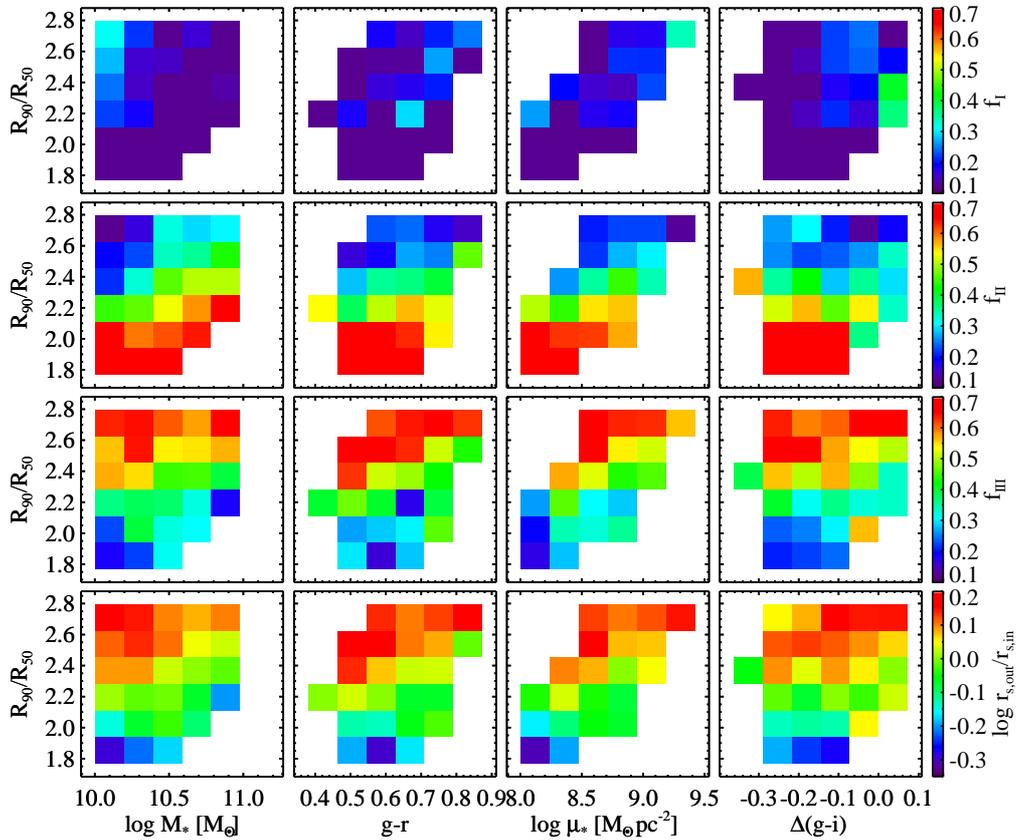}
\caption{The fractions of galaxies hosting type I, II and III discs as a function of other galactic properties. From top to bottom, the panels are color coded by the fractions of type I, II and III galaxies ($\fI$, $\fII$ and $\fIII$). The range of the parameter space in each dimension is determined by 1 and 99 percentile of the parameter distribution of the main sample, and is evenly divided into 6 bins (pixels). Pixels with less than 20 galaxies are set to be blank (white).  }
\label{fig:typefrac}
\end{figure*}

In the top three rows of Figure~\ref{fig:typefrac}, among the parameters investigated the most prominent trends of type fractions and $r_{out}/r_{in}$ are with $R_{90}/R_{50}$. $\fII$ rises from $\sim20\%$ at $R_{90}/R_{50}=2.7$ to $\sim80\%$ at $R_{90}/R_{50}=2.0$ (second row of Figure~\ref{fig:typefrac}), while $\fIII$ decreases from $\sim70\%$ to $\sim10\%$ across the same $R_{90}/R_{50}$ range (third row of Figure~\ref{fig:typefrac}). $\fI$ is low throughout the parameter space ($<25\%$). When $R_{90}/R_{50}>2.5$ there is a weak trend for more massive galaxies to have slightly higher $\fII$ and lower $\fIII$ than the less massive galaxies. $\fII$ and $\fIII$ change insignificantly as a function of $\mu_*$, $g-r$ or $\Delta(g-i)$.

In the bottom row of Figure~\ref{fig:typefrac}, the scale-length ratio $r_{s,out}/r_{s,in}$ correlates well with $R_{90}/R_{50}$, and weakly with $M_*$ when $R_{90}/R_{50}>2.3$. The trend is consistent with our discussion for $\fI$, $\fII$ and $\fIII$. $r_{s,out}/r_{s,in}$ fully parametrize the three different disc types. However, in the following we will discuss relatively weak trends which can only reveal after we control for the $R_{90}/R_{50}$ and $M_*$. The relatively small sample size only allows us to discuss the different behavior of galaxies with a relatively small or large value of $r_{s,out}/r_{s,in}$. The classification of type II and III is a natural division of the sample for such a purpose. Therefore, in the following sections, we only use the parameters $\fI$, $\fII$ and $\fIII$, instead of $r_{s,out}/r_{s,in}$ to study the distribution of type I, II and III discs among different populations of galaxies.

To summarize, $R_{90}/R_{50}$ and $M_*$ are identified to be the major parameters to be controlled in our further analysis of $\fIII$ dependencies on other properties.

\subsection{The dependence of $\fIII$ on H{\textsc I}-richness} 
\label{sec:HIprop}
Because the in-situ star formation is presumably an expected way of adding stars to the outskirts of discs, and because the star formation in the outer discs is closely related with the H{\textsc I}-richness \citep{Wang11}, we investigate the possible relation between $\fIII$ and $\Delta f_{\rm HI}$ (a measure of the H{\textsc I}-richness, see Section~\ref{sec:HIdata}) in this section.

\begin{figure*} 
\includegraphics[width=14cm]{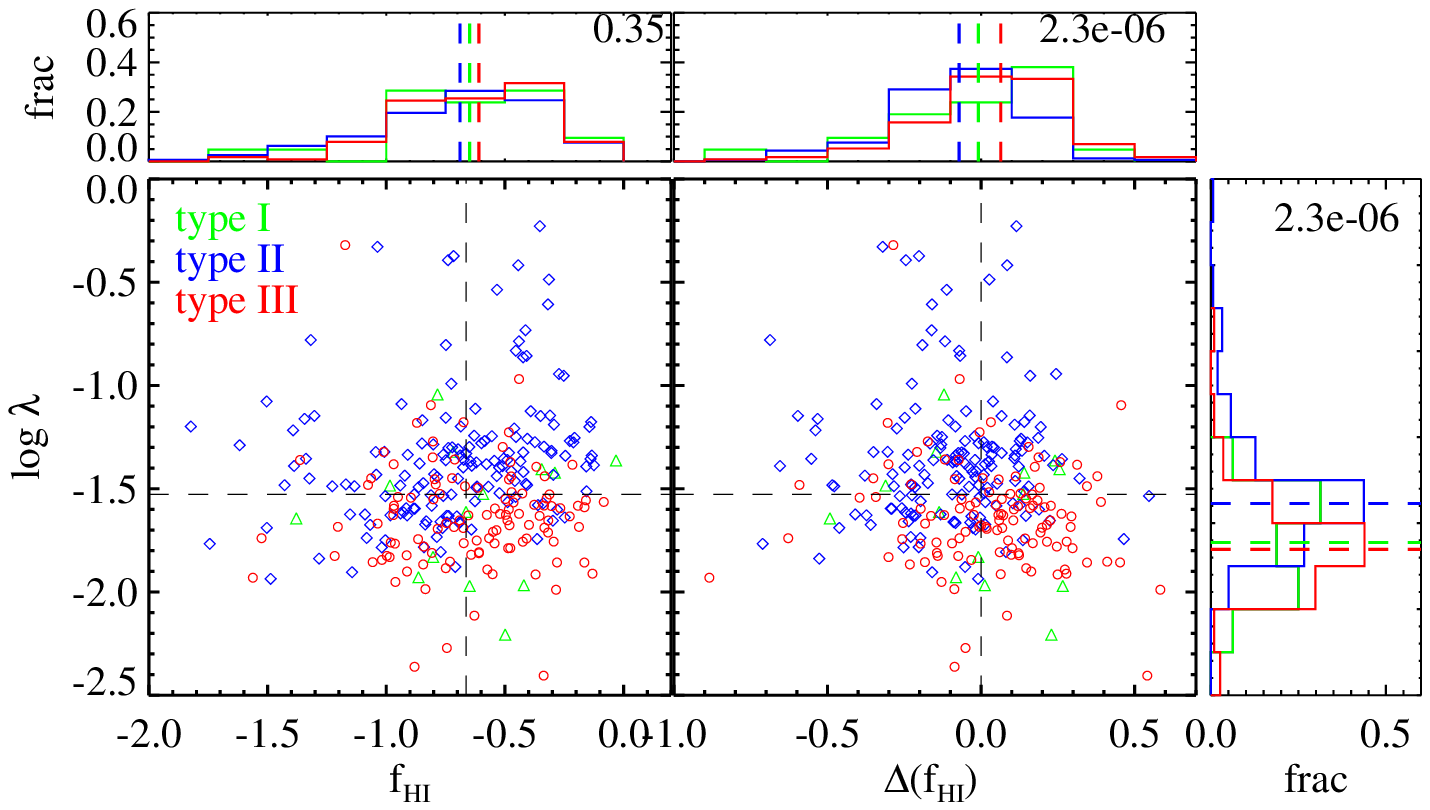}
\caption{The distribution of type I, II and III galaxies in the parameter space of f$_{\rm HI}$, $\Delta f_{\rm HI}$ and $\lambda$. Type I, II and III galaxies are plotted in green, blue and red respectively.  The dashed lines show the median value of the histograms. In each panel that shows histograms, we denote in the corner the K-S test probability which indicates the similarity between the type II and III sub-samples in the property distribution.}
\label{fig:property2d_HI}
\end{figure*}

Figure~\ref{fig:property2d_HI} shows that type III galaxies tend to have lower $\lambda$, higher $\Delta f_{\rm HI}$ and slightly higher f$_{\rm HI}$ than type II galaxies. $\sim$75\% of the H{\textsc I}-rich ($\Delta f_{\rm HI}>0$) and low-$\lambda$ galaxies are type III galaxies. The question is whether this reflects an intrinsic correlation of type III breaks preferring H{\textsc I}-rich and low $\lambda$ galaxies or by these parameters depending on $R_{90}/R_{50}$ or $M_*$ of galaxies. 

Therefore, we make a $R_{90}/R_{50}$-control sample for each special sub-sample under investigation. The control sample is built by randomly selecting galaxies from the main sample and matching the sub-sample in the normalized distributions of $R_{90}/R_{50}$ and $\log M_*$. The bin size for matching distributions is 0.1 for both $R_{90}/R_{50}$ and $\log M_*$. We ensure the goodness of matching by requiring the K-S (Kolmogorov-Smirnov) test probabilities to be above 0.85 when comparing the distributions of parameters between the two samples. The final size of the control sample can be as large as or up to ten times larger than the sub-sample under investigation.

In the top row of Figure~\ref{fig:typefrac_HI}, the H{\textsc I} sample and its $R_{90}/R_{50}$-control sample (purple and orange bars) have similar type I-III fractions within the error bar, lending support to our method of controlling the $R_{90}/R_{50}$ and $M_*$ parameters. The H{\textsc I}-rich sample has noticeably higher $\fIII$ than its $R_{90}/R_{50}$-control sample (cyan and pink bars). In the first column of Figure~\ref{fig:colorprofile_type3}, we compare the median $g-r$ colour profile of the H{\textsc I}-rich type III galaxies with that of control galaxies which are selected from the main sample and matched in total $g-r$ and $M_*$ ($(g-r)$-control sample hereafter). The former is bluer by $\sim0.03$ mag (with a 1.5-2$\sigma$ significance), indicating $\sim$1.4 times higher SFR$/M_*$ \citep[assuming the solar abundance and a constant star forming history][]{BC03, Kauffmann03}, than the latter in the outer region. These results indicate that the H{\textsc I}-rich type III galaxies have more actively growing outer disc due to the star formation. 

\subsection{The dependence of $\fIII$ on the spin parameter}
Simulations predict that galaxies formed in low-spin halos tend to form type III breaks via bar-driven stellar migration  \citet{Herpich15}, so we investigate the relation between $\fIII$ and $\lambda$ in this section.

\begin{figure} 
\includegraphics[width=9cm]{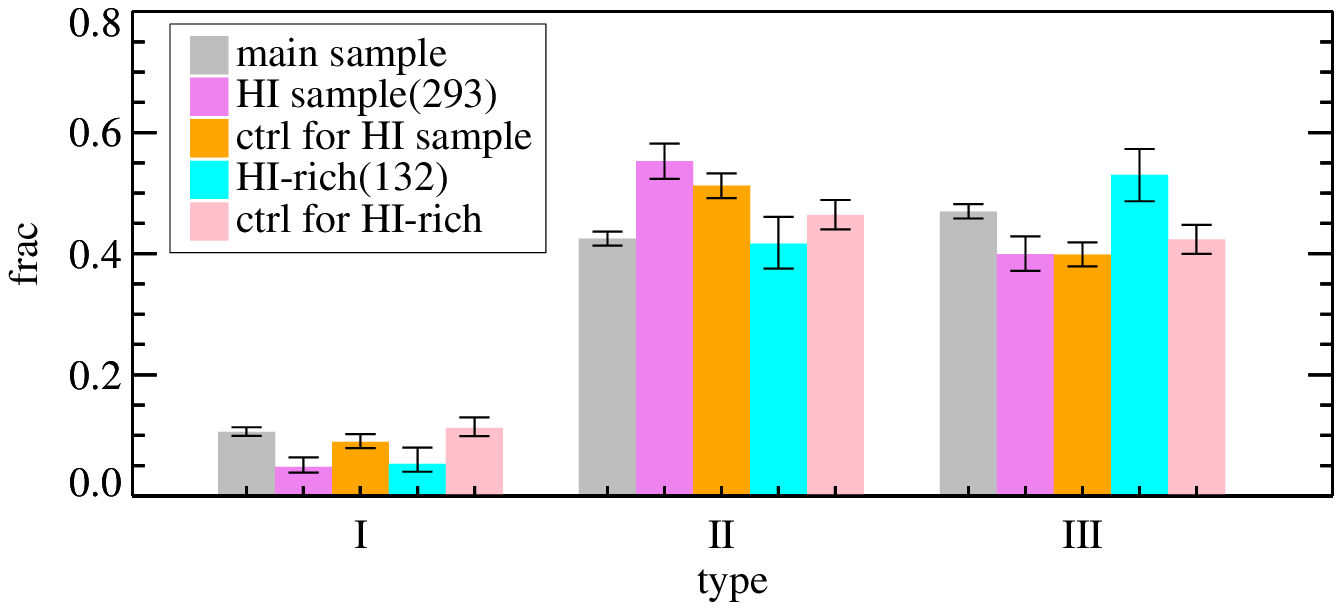}

\includegraphics[width=9cm]{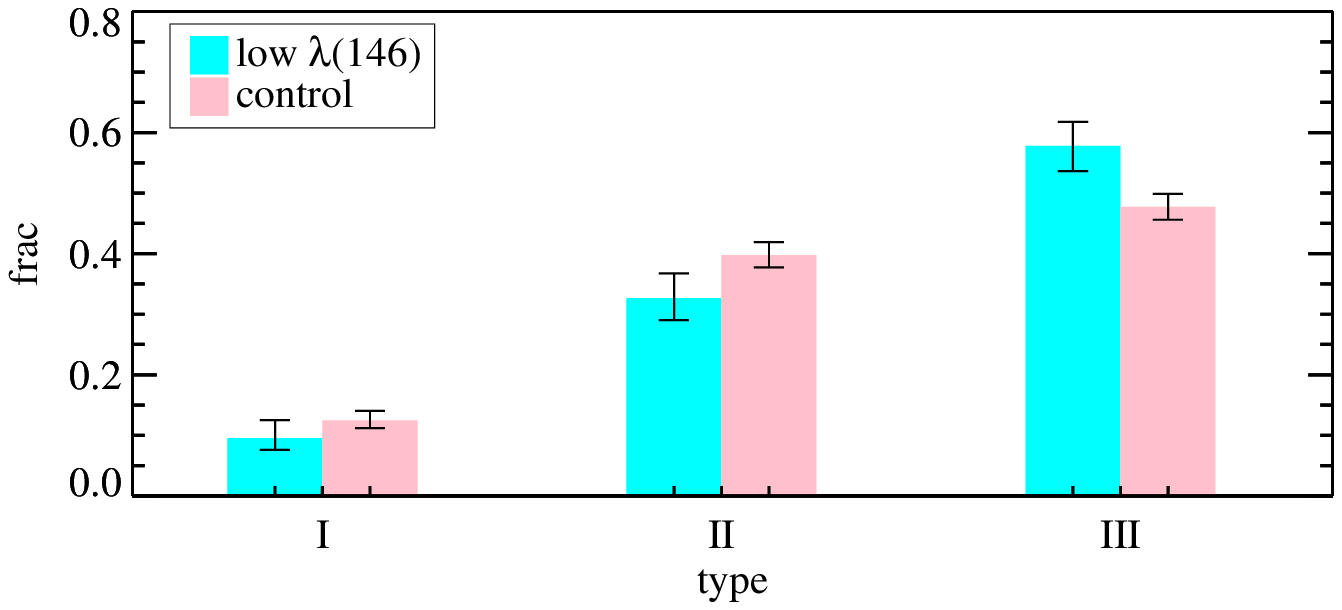}

\includegraphics[width=9cm]{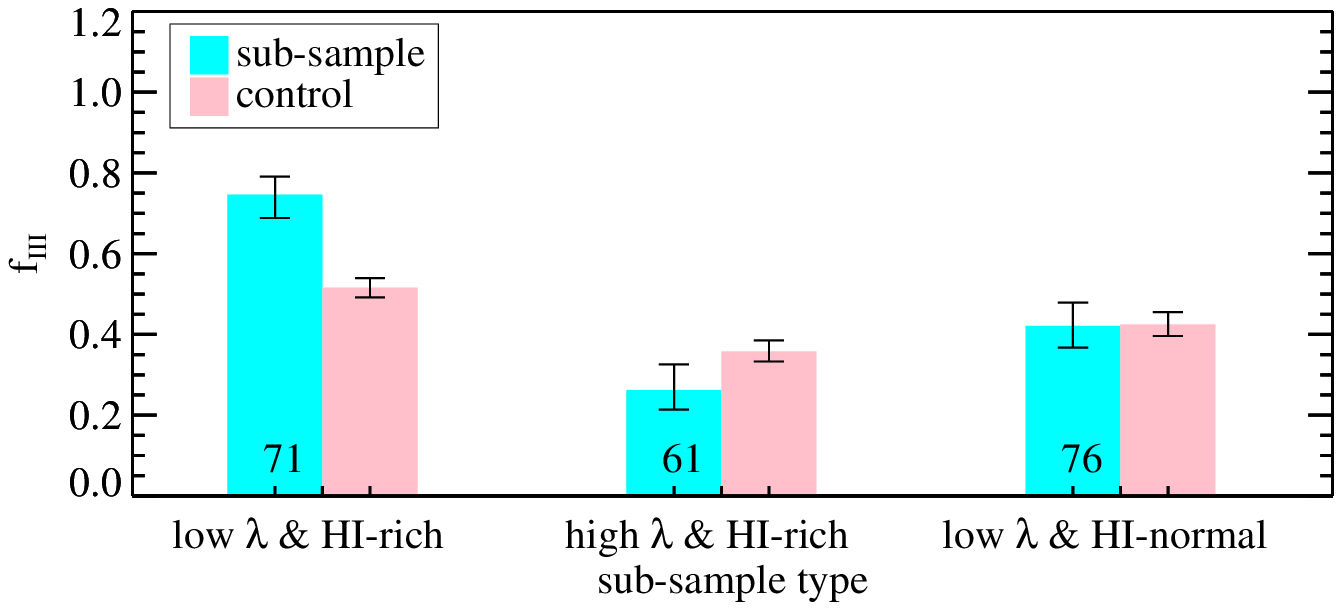}

\caption{Bar graph for the disk break fractions. In the top and middle panels, the three groups of bars are for $\fI$, $\fII$ and $\fIII$ respectively; the numbers in the brackets denote the number of galaxies in the sub-samples. In the bottom panel, the three groups of bars are for $\fIII$ of different sub-samples and their corresponding control samples (see text in Section~\ref{sec:HIprop}); the numbers near the bottom of bars denote the number of galaxies in the sub-samples. The colors representing different samples are denoted in the top-left corner. ``ctrl'' and ``control''  refer to the $R_{90}/R_{50}$-control samples. The error bars are 68.3\% (1-$\sigma$ for a gaussian distribution) confidence intervals calculated for a binomial population \citep{Cameron11}. }
\label{fig:typefrac_HI}
\end{figure}

\begin{figure*} 
\includegraphics[width=5.5cm]{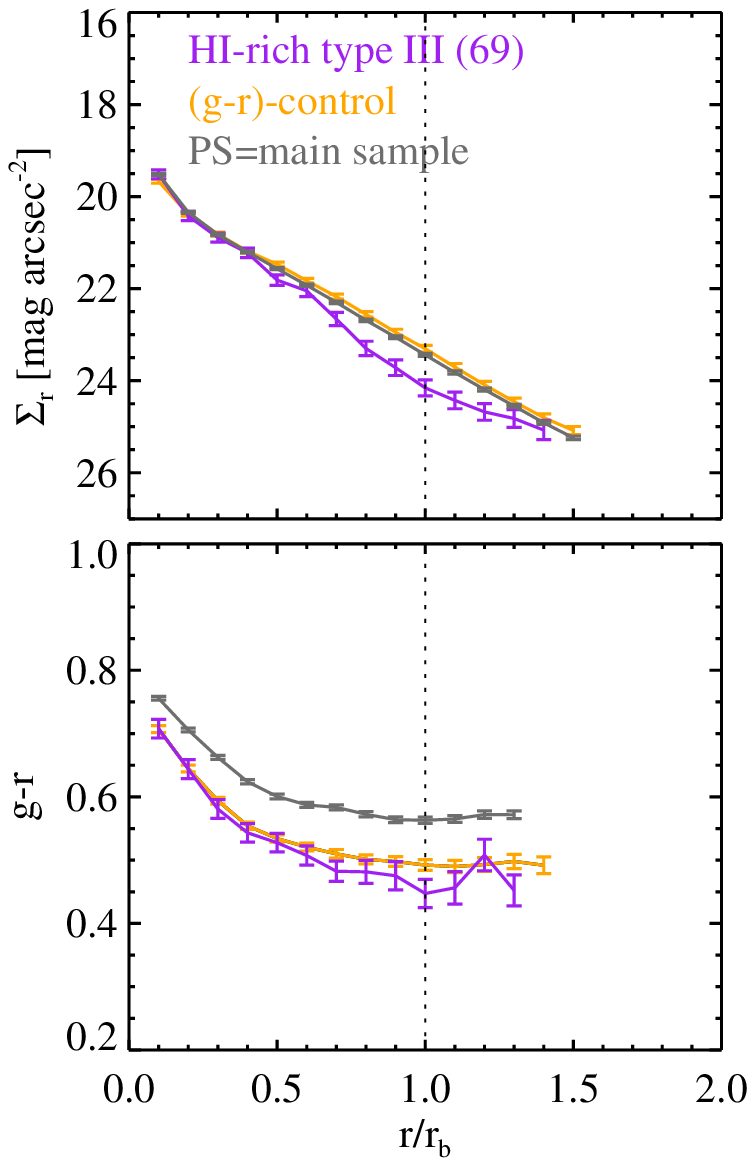}
\includegraphics[width=5.5cm]{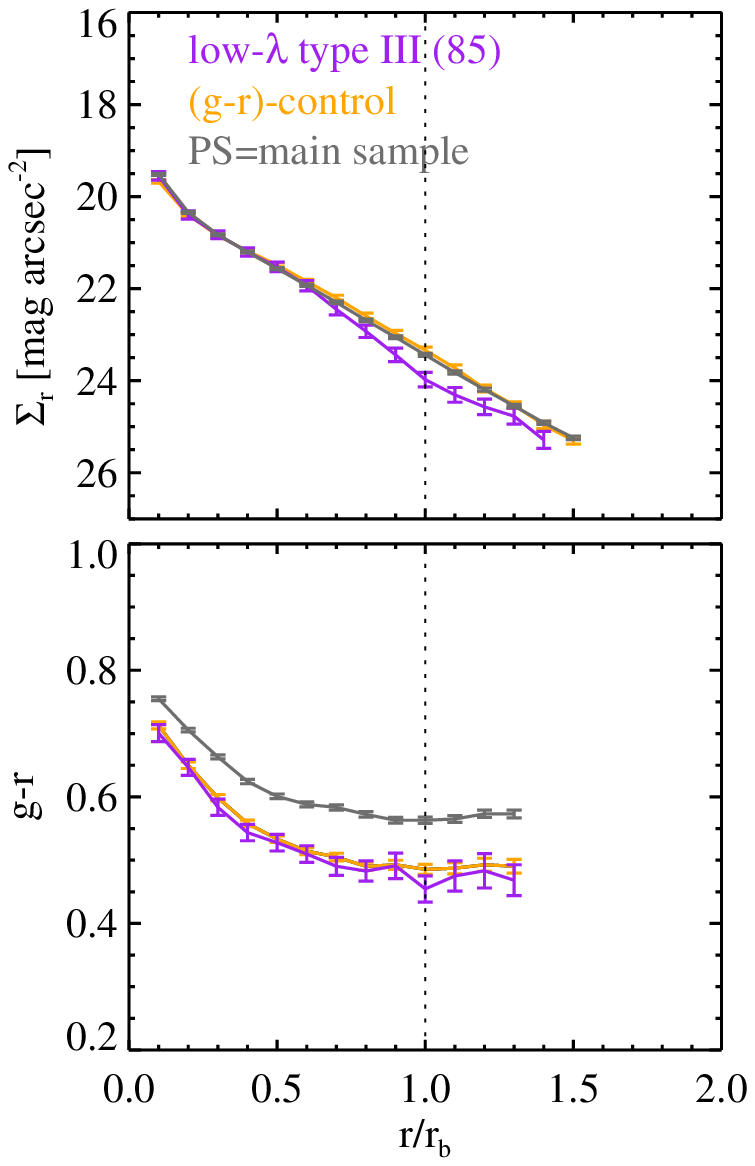}
\includegraphics[width=5.5cm]{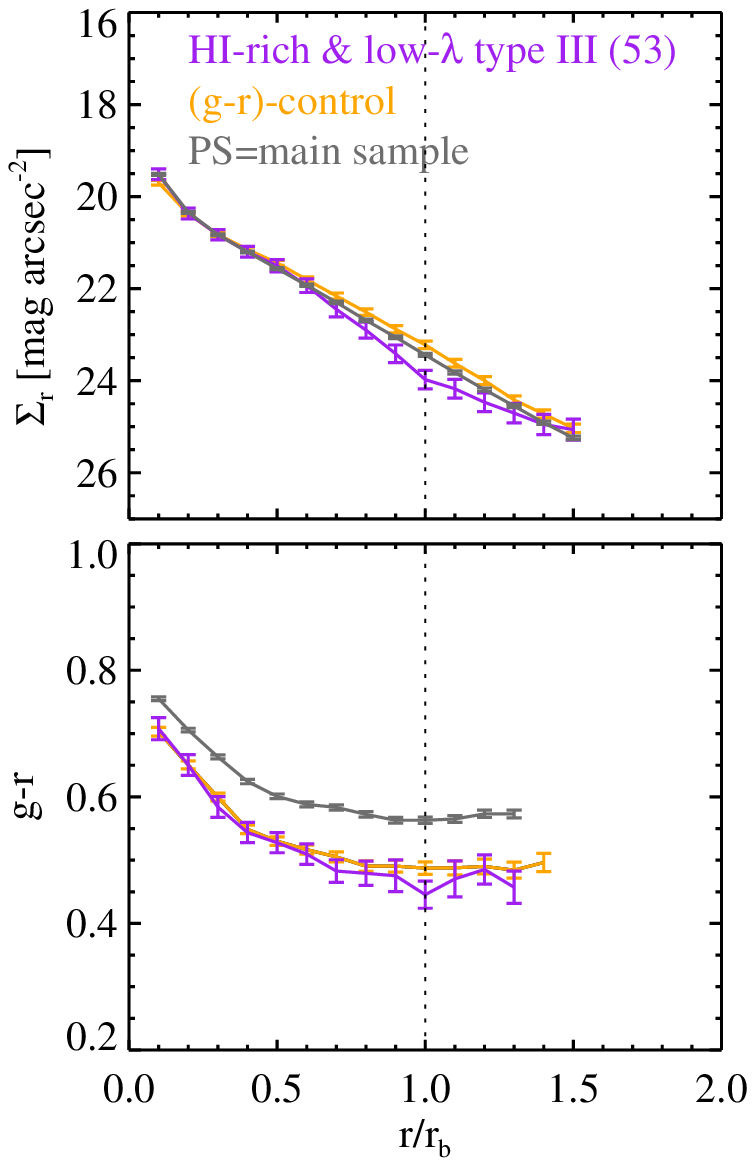}
\caption{The median surface brightness and color profiles. It compares sub-samples of type III galaxies with control galaxies that are selected from the main sample and matched in the global $g-r$ colour and $M_*$ (($g-r$)-control samples hereafter). The three columns are for sub-samples of type III galaxies that are H{\textsc I}-rich (left), low in $\lambda$ (middle), and both H{\textsc I}-rich and low in $\lambda$ (right), respectively. Purple curves are for type III galaxies, orange curves are for the ($g-r$)-control samples, and black curves are for all galaxies in the main sample. 
 Error bars show the deviation of the median values. } %
\label{fig:colorprofile_type3}
\end{figure*}

In the middle row of Figure~\ref{fig:typefrac_HI}, the low-$\lambda$ sample also has higher $\fIII$ than its $R_{90}/R_{50}$-control sample. The median colours of low-$\lambda$  type III galaxies do not differ much from their ($g-r$)-control galaxies (middle column of Figure~\ref{fig:colorprofile_type3}). Hence the enhanced type III break formation in low-$\lambda$ galaxies is not related to enhanced star formation in the outer discs.

To this point, we ask, what are the relative importances of H{\textsc I}-richness and $\lambda$ in regulating $\fIII$ of galaxies. The small sample size does not allow us to further control H{\textsc I}-richness or $\lambda$. Instead, we search for clues by investigating the $\fIII$ of sub-samples with high$\slash$low H{\textsc I}-richness and high$\slash$low $\lambda$. In the bottom row of Figure~\ref{fig:typefrac_HI}, we find that the H{\textsc I}-rich but high $\lambda$ galaxies, and the low-$\lambda$ but H{\textsc I}-poor galaxies both show lower $\fIII$ than the $R_{90}/R_{50}$-control galaxies. The sub-sample with both high H{\textsc I}-richness and low $\lambda$ shows a significant (higher than $3\sigma$) enhancement in $\fIII$ when compared to its $R_{90}/R_{50}$-control sample. These galaxies have a surprisingly high $\fIII\sim75\%$. 
From the right column of Figure~\ref{fig:colorprofile_type3}, we can see that their median $g-r$ profile is tentatively bluer than its ($g-r$)-control sample in the outer region (right column of Figure~\ref{fig:colorprofile_type3}), and in this way they are like an intermediate population between the H{\textsc I}-rich galaxies and the low-$\lambda$ galaxies.
These results imply that the effects of star formation driven by gas accretion and stellar migration enhanced in low-spin halos may be both important in the H{\textsc I} sample of this paper.

\subsection{The dependence of $\fIII$ on the existence of strong bars}
\label{sec:bar}
Simulations suggest that bars strongly promote the formation of type III breaks via scattering stars to elongated orbits \citep{Herpich17}. We examine such an effect and check how it is related with the effects of $\Delta f_{\rm HI}$ and $\lambda$.

\begin{figure} 
\includegraphics[width=9cm]{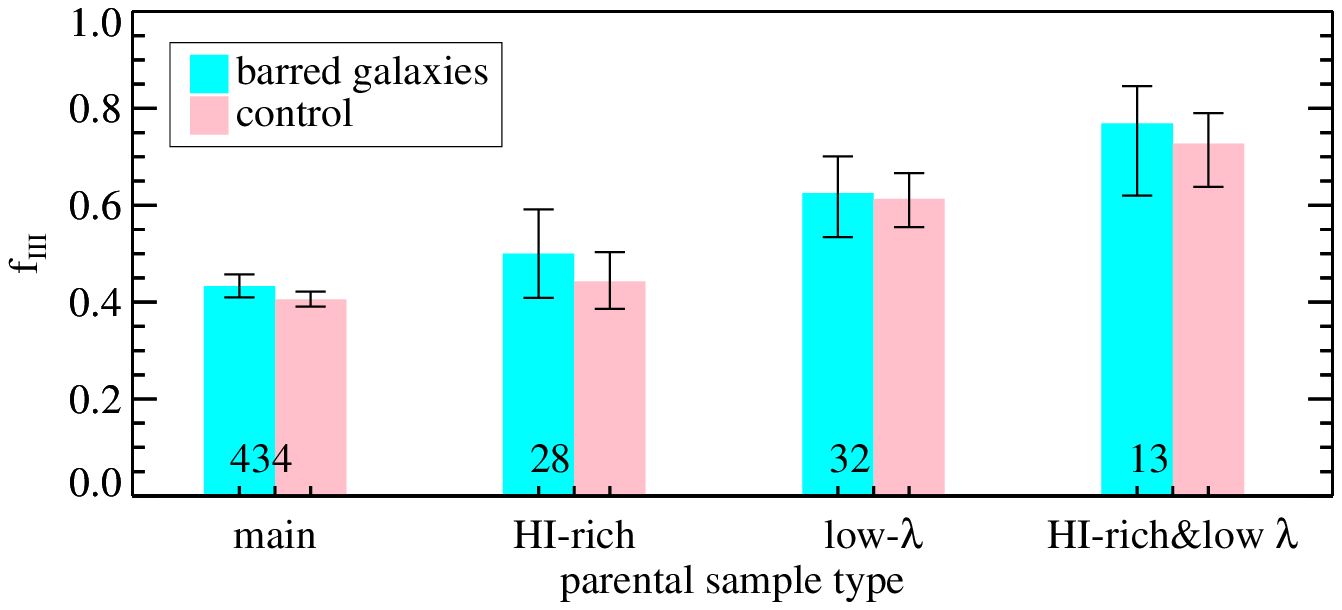}

\includegraphics[width=9cm]{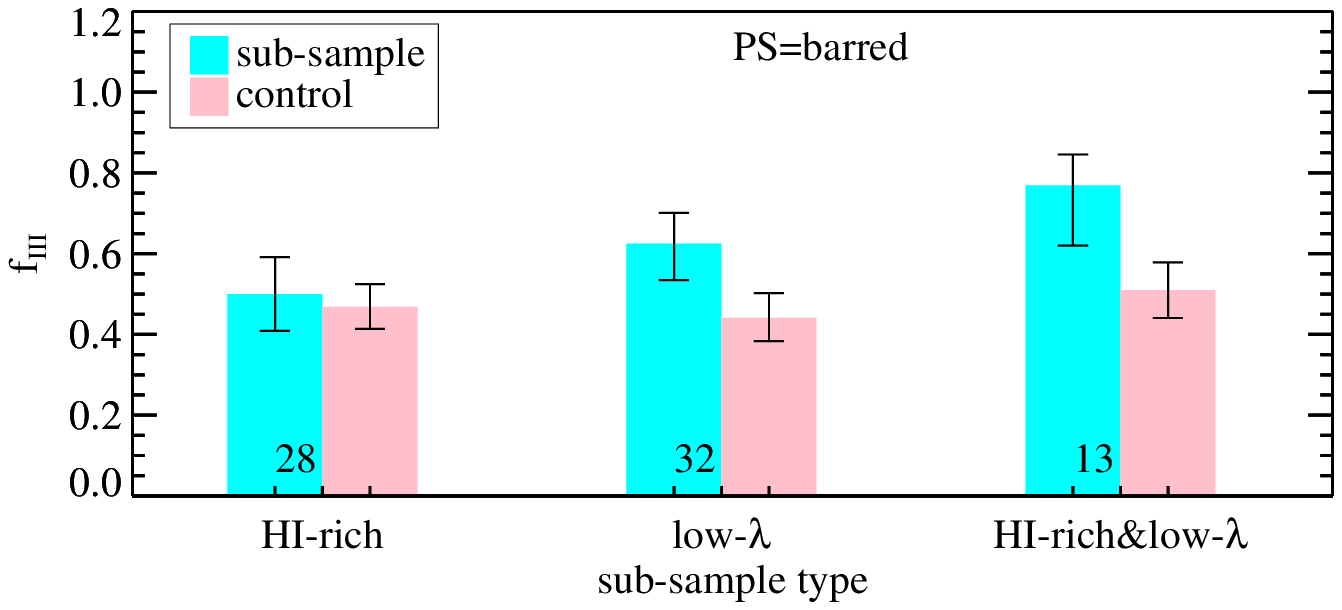}

\caption{Bar graph for the type III fractions of different samples. Top: it compares barred galaxies with galaxies that are not necessarily barred. Each cyan bar is for a sub-sample of barred galaxies, and the closest pink bar is for its $R_{90}/R_{50}$-controlled sample. The x-axis denotes the selection of parent samples from which both the barred and $R_{90}/R_{50}$-controlled sub-sample are selected. Bottom: it compares special barred galaxies which are selected in H{\textsc I}-richness or $\lambda$ with barred galaxies which are not selected in H{\textsc I}-richness or $\lambda$. The parent sample is the barred galaxies from the main sample; all sub-samples and control samples in this panel are selected from this parent sample. Cyan bars are for a sub-samples under investigation, with the name denoted on the x-axis, and the pink bars are for the corresponding $R_{90}/R_{50}$-controlled samples. The numbers near the bottom of bars denote the number of galaxies in the sub-samples}
\label{fig:typefrac_bar}
\end{figure}

In the top row of Figure~\ref{fig:typefrac_bar}, we take four parent samples: the main sample and three H{\textsc I} sub-samples. In each parent sample, we compare $\fIII$ of the barred galaxies with the $R_{90}/R_{50}$-control sample of galaxies that are not necessarily barred \footnote{Our results do not significantly change if the control samples are selected from the unbarred galaxies.}. We find that the existence of strong bars is not related with significantly enhanced $\fIII$ with respect to the $R_{90}/R_{50}$-control galaxies in any of these parent samples. This result seems to be at odd with the theoretically predicted scattering effects of bars \citep{Herpich15}, and we will discuss it with more details in Section~\ref{sec:discussion}.

How is $\fIII$ correlated with H{\textsc I}-richness and $\lambda$ in barred galaxies? In the bottom row of Figure~\ref{fig:typefrac_bar}, we compare $\fIII$ of the barred galaxies 
which are selected in H{\textsc I}-richness and$\slash$or $\lambda$ with the barred galaxies which are not selected in H{\textsc I}-richness or $\lambda$. We find that barred galaxies which are both H{\textsc I}-rich and low-$\lambda$ have higher $\fIII$ than their $R_{90}/R_{50}$-control sample of barred galaxies (with a 1.5$\sigma$ difference). Hence the enhanced $\fIII$ in H{\textsc I}-rich and low-$\lambda$ galaxies is independent of the existence of a strong bar. 

Galaxies which are only low in $\lambda$ (but without a selection in H{\textsc I}-richness) also have higher $\fIII$ than the $R_{90}/R_{50}$-control galaxies.
On the other hand, for galaxies which are only H{\textsc I}-rich (but without a selection in $\lambda$), there is no difference in $\fIII$ when comparing to their $R_{90}/R_{50}$-control sample. 

\subsection{The dependence of $\fIII$ on environment}
\label{sec:env}
Galactic environments may play a role in forming type III breaks. Stellar accretion through mergers may directly add stars to the outer discs \citep{Cooper13}. Simulations predict that minor mergers can also effectively form type III breaks by inducing strong gas inflow and causing the outer disc to expand \citep{Younger07}. These effects may be closely related to our main results presented in Section~\ref{sec:HIprop}, because in simulations the merger induced stellar migration is more prominent in H{\textsc I}-rich galaxies than in H{\textsc I}-poor galaxies \citep{Younger07}, and gas-rich minor mergers tend to increase the $\lambda$ of galaxies \citep{Lagos18}. Moreover, analysis based stacked H{\textsc I} spectrum suggests that satellites in low-mass groups are significantly more H{\textsc I}-rich than satellites in massive groups \citep{Brown17}.

We are unable to directly examine the merger history or reliably identify minor merger remnants with the relatively shallow SDSS images used in this paper. However, we can use the environment sample built upon the Group Catalogue \citep{Lim17} to gain some clues.

\begin{figure} 
\includegraphics[width=9cm]{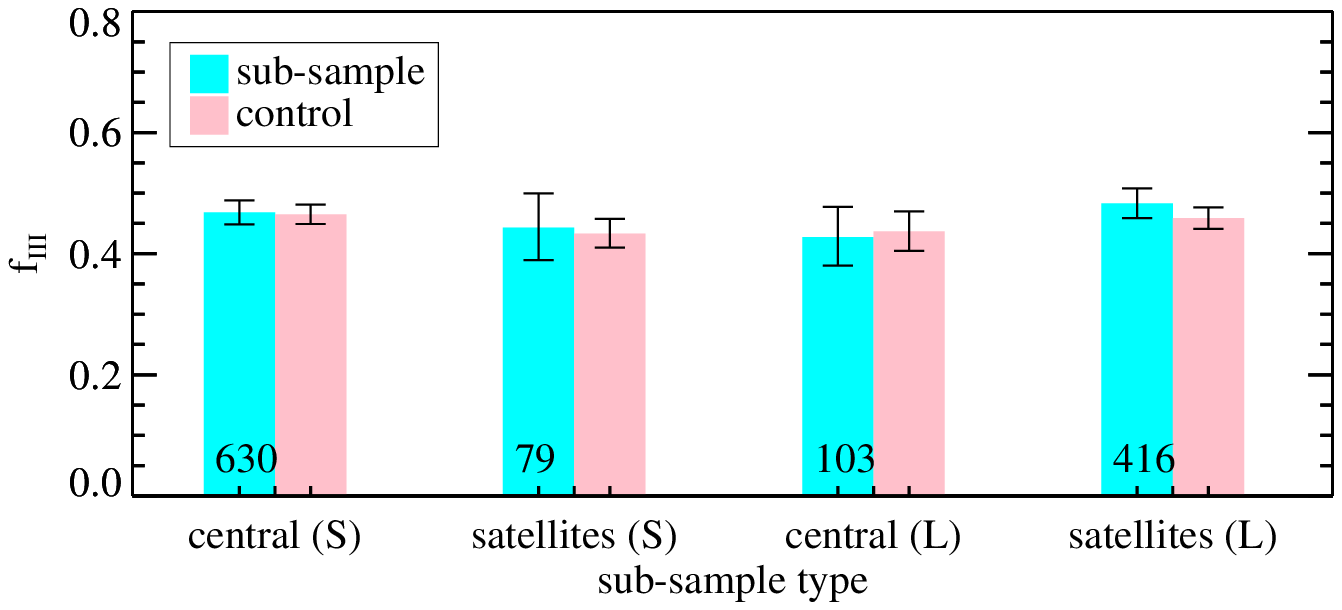}
\caption{Bar graph for the type III fractions of central and satellite galaxies. The cyan color is for the sub-samples under investigation and the pink color is for the $R_{90}/R_{50}$-control samples selected from the main sample.  ``S'' stands for small groups with $M_{group}<12.5~M_{\odot}$, and ``L'' stands for large groups with $M_{group}>12.5~M_{\odot}$. The numbers near the bottom of bars denote the number of galaxies in the sub-samples. }
\label{fig:typefrac_central}
\end{figure}

We expect the merger history to be richer for central galaxies than for satellites, and richer for central galaxies of more massive groups than for those of less massive groups \citep{Blanton09}. Therefore we divide the environment sample into four sub-samples of galaxies: central$\slash$satellite galaxies of small$\slash$large groups ($M_{\rm group}$ lower $\slash$higher than $12.5~M_{\odot}$), and compare them with $R_{90}/R_{50}$-control samples which are not selected in environmental properties. Figure~\ref{fig:typefrac_central} shows that the $\fIII$ of each sub-sample is not significantly higher than the corresponding $R_{90}/R_{50}$-control sample. 

\begin{figure} 
\includegraphics[width=9cm]{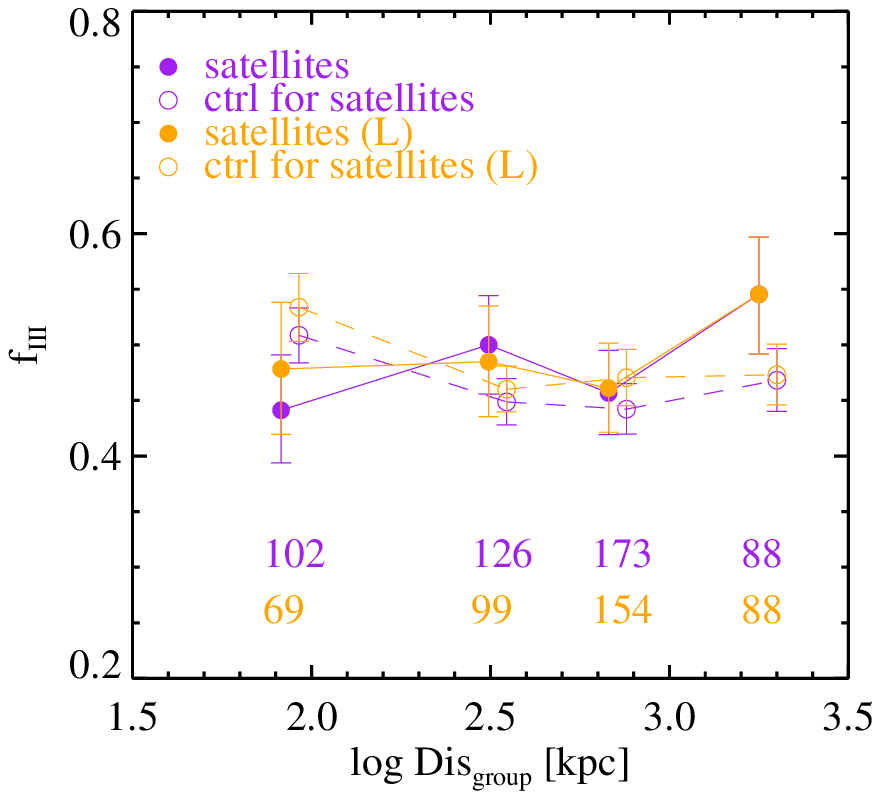}
\caption{Type III fractions of satellite galaxies as a function of distance from the group center. The colours and symbols for different sub-samples are denoted in the top-left corner. ''ctrl'' stands for $R_{90}/R_{50}$-control samples and ``L'' stands for large groups with $M_{group}>12.5~M_{\odot}$. The number of galaxies in each sub-samples is denoted near the bottom of the figure. }
\label{fig:typefrac_sat}
\end{figure}

Because the satellite galaxies at different distances from the group centers vary greatly in their properties \citep{Blanton09}, we further divide the satellites into sub-samples by their distance to group centers (Figure~\ref{fig:typefrac_sat}). Again we do not find them to have significantly higher $\fIII$ than their $R_{90}/R_{50}$-control samples. 

We perform a similar analysis as in the bottom panel of Figure~\ref{fig:typefrac_bar}, to test whether $\fIII$ of the central or satellite galaxies have a different dependence on H{\textsc I}-richness or $\lambda$. We do not find any significantly different trends when comparing the sub-samples of central or satellite galaxies.

Hence we find no correlation between $\fIII$ and the environmental properties investigated here. The result is consistent with previous observational studies \citep{Laine14, Maltby12}. 

\section{Discussion}
\label{sec:discussion}
We have studied the fraction of galaxies hosting type III breaks (identified in the optical $r$-band) for a main sample of $\sim$2000 galaxies from SDSS and a sub-sample of $\sim$300 galaxies with H{\textsc I} single-dish measurements from ALFALFA.  \citet{Huang12} showed that galaxies in H{\textsc I} samples as used in this paper (from cross-matching ALFALFA, SDSS and GALEX) well cover the SFMS (star forming main sequence). So the underlying physics indicated with this sub-sample of H{\textsc I}-rich and low-$\lambda$ galaxies are of statistic importance for the formation of massive star-forming disc galaxies. We summarize our key results below.

\begin{enumerate}
\item The fraction of galaxies hosting type III breaks ($\fIII$) most strongly depends on $R_{90}/R_{50}$ (higher $\fIII$ for higher $R_{90}/R_{50}$, consistent with the result of Z15), and weakly depends on $M_*$ (higher $\fIII$ for lower $M_*$) (Section~\ref{sec:optprop}).

\item H{\textsc I}-rich and low-$\lambda$ galaxies have a high $\fIII$ of $\sim75\%$, which is significantly higher than the  $R_{90}/R_{50}$-control (controlled for both $R_{90}/R_{50}$ and $M_*$) galaxies. The trend is significant in strongly barred galaxies, and is regardless of the environment.

\item  H{\textsc I}-rich galaxies have higher $\fIII$ than $R_{90}/R_{50}$-control galaxies. The trend is insignificant in strongly barred galaxies.

\item Galaxies with low $\lambda$ have higher $\fIII$ than $R_{90}/R_{50}$-control galaxies. 

\item Galaxies with strong bars do not have significantly higher $\fIII$ than $R_{90}/R_{50}$-control galaxies.

\item The identity of galaxies as central or satellite galaxies, the halo mass of the groups where the galaxies reside, or the distance of satellite galaxies to the group centers hardly enhance $\fIII$ of galaxies. 

\end{enumerate}

The dependence of break types on $R_{90}/R_{50}$ demonstrates that type II and III galaxies do not just differ in the outskirts, but also in the inner discs. It suggests a possible physical coupling in the formation of the inner and outer discs. A detailed discussion on this topic would rely on a careful bulge-disc decomposition for the inner parts of the galaxies, and is beyond the scope of this paper. In control of $R_{90}/R_{50}$, our results point to two major conditions that are related to the formation of type III breaks: high H{\textsc I}-richness and low $\lambda$. We highlight the surprisingly $\fIII$ ($\sim75\%$) in the H{\textsc I}-rich and low-$\lambda$ galaxies, which seems to be independent of $R_{90}/R_{50}$, the existence of strong bars and environment. This sub-sample of galaxies consists of 26.9\% of the galaxies, but host 55\% of the type III breaks from the whole H{\textsc I} sample. The high H{\textsc I}-richness may be related to the gas accretion and inside-out disc formation \citep{Mo98, Minchev12}, while the process responsible for the relation between low $\lambda$ and $\fIII$ is less clear. 

\begin{figure} 
\includegraphics[width=8cm]{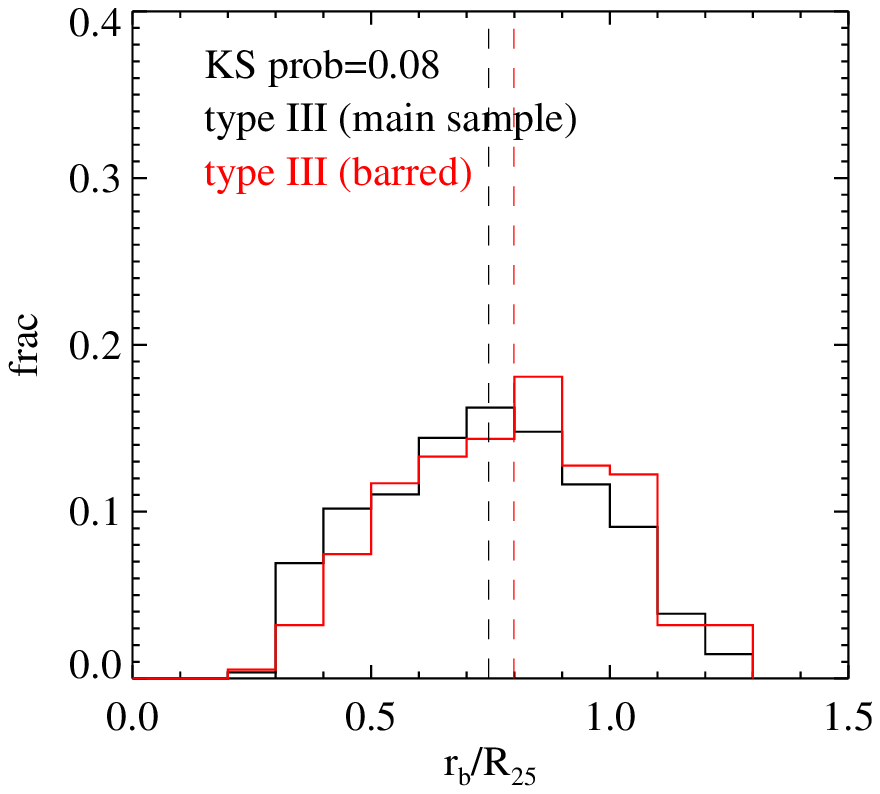}
\caption{Distribution of $r_b/R_{25}$ for type III galaxies. The black and red lines are for type III galaxies from the main sample and the barred sample respectively. The median values are marked with dashed lines. The KS-test probability indicating the similarity of the two distributions are denoted. }
\label{fig:rb_type3}
\end{figure}

The previous studies suggest that strong bars, low-$\lambda$ discs and type III discs are closely related. Numerical simulations predict that galaxies with low $\lambda$ are more likely to develop global instability and form strong bars \citep{Efstathiou82, Foyle08}, which is supported by previous observations that long bars tend to exist in galaxies with low $\lambda$ \citep{CervantesSodi13}. Simulations also predict that in low-$\lambda$ discs, strong bars strongly scatter inner stars initially on circular orbits to highly elongated orbits with large major axis and consequently form type III breaks \citep{Herpich15, Herpich17}. 
These results suggest a picture where a low-$\lambda$ galaxy tend to form a type III disc via the formation and scattering effect of a strong bar. Hence it is a bit surprising to find in this paper that the relation between low $\lambda$ and the formation of type III discs seems to be independent from the existence of strong bars. Previous theoretical studies have shown that one possibility is that strong bars that have already dissipated may have facilitated the formation of type III discs, as well as the growth of the central concentrations responsible for the dissipation of the strong bars \citep[Figure 7 in ][]{Herpich17}. 
Another possibility is that the type III breaks produced by strong bars were not identified in this paper. If most of the stars scattered by bars move to radius larger than $r_{out}$, then the depth of our data will not allow us to observe the related disc breaks. We find that on average the barred type III galaxies have larger $r_b/R_{25}$ than the type III galaxies from the main sample (Figure~\ref{fig:rb_type3}), which seems to support this speculation. 
Finally, low $\lambda$ and type III breaks may be both connected to the same galaxy key properties but not in a causal fashion. If so, more observables than those used here will be needed to identify those key properties in the future.

\begin{figure} 
\includegraphics[width=8cm]{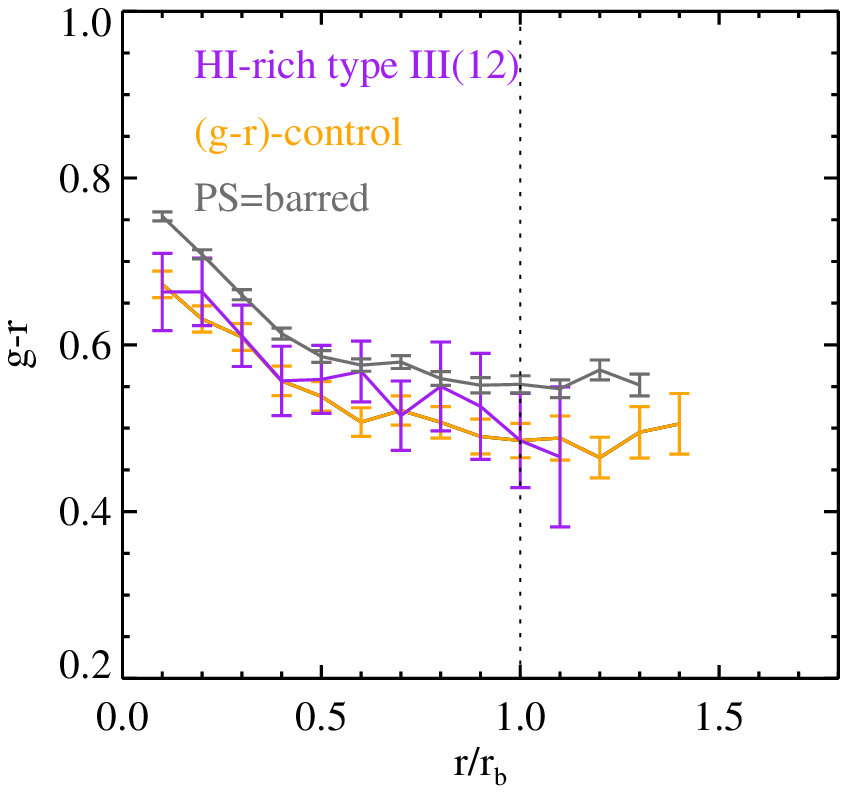}
\caption{The median color profiles. Similar as the bottom-left panel of Figure~\ref{fig:colorprofile_type3}, but the sub-sample under investigation is changed to H{\textsc I}-rich barred galaxies.}
\label{fig:colorprofile_bar}
\end{figure}

It is not direct from our result whether strong bars are related to the effect of low $\lambda$ in forming type III disc breaks, but the strong bars seem to weaken the effect of forming type III breaks via gas accretion and star formation, as we find that H{\textsc I}-rich barred galaxies do not show higher $\fIII$ than the $R_{90}/R_{50}$-control barred galaxies. The weakening effect is possibly due to the strong inflow of gas driven by the torques of strong bars \citep{Schwarz81, Wang12}, which quickly removes the gas from the outer discs and reduce the star formation rate there. If so, we would expect H{\textsc I}-rich barred galaxies to not be significantly bluer in their outer discs than the $(g-r)$-control barred galaxies, contrary to what we have found in Figure~\ref{fig:colorprofile_type3} for H{\textsc I}-rich galaxies that are not selected in the bar properties. Figure~\ref{fig:colorprofile_bar} tentatively supports this scenario, but is limited by the large error bars due to the small number of H{\textsc I}-rich barred type III galaxies (in total 12). A larger sample will be needed to confirm the result in the future. 

The effect of stellar accretion and migration caused by mergers might be weak or be counteracting with other processes.
Hydrodynamic simulations predict that galactic interactions during mergers induce strong inflow of the cold gas and may trigger an inner starburst \citep{Barnes92}. As a result, the star formation may get enhanced in the galactic center and suppressed in the galactic outskirts. Hydrodynamic simulations predict that wet mergers (gas-rich) tend to increase $\lambda$ of galaxies while dry mergers (gas-poor) tend to decrease them \citep{Lagos18}.  
As a result, wet mergers may destroy the internal condition prone to the formation of  type III beaks.  
So although minor mergers in gas-rich galaxies can trigger the formation of type III breaks in some simulations \citep{Younger07}, they may weaken the effect of the gas accretion and processes related to low $\lambda$, and thereby undermine the net effect in forming the type III breaks. These speculations may partly explain why no significant correlation between type III discs and the environment has been identified with the relatively small sample of this paper (Section~\ref{sec:env}). 
The real interplay between these different processes needs to be coherently studied in simulations and larger observational datasets in the future.

Finally, we warn that some of the analysis and discussion about the interplay between H{\textsc I} properties, strong bars and environment in regulating $\fIII$ are based on relatively small sub-samples of a few tens of galaxies, and should be taken with caution. We also remind the reader that many broken discs have more than one break (28.7\%, 34.9\%, 29.7\% and 23.2\% for the main sample, the barred galaxies, the central galaxies and the satellite galaxies respectively), and we have selected the outmost breaks for investigation in this paper. Removing the galaxies with multiple breaks does not significantly change the trends studied in this paper. These more complex systems may provide clues on the combined effects of different mechanisms, and will be investigated in the future.

\section{Conclusion and future perspective}
\label{sec:conclusion}
The most important finding of this paper is that high H{\textsc I}-richness and low $\lambda$ of the inner stellar discs are related to an enhanced formation of type III discs. This is consistent with numerical models where gas accretion brings raw material for forming stars in the outer discs. It remains unclear whether bar-driven stellar migration is responsible for the correlation between low $\lambda$ and the formation of type III discs. 

Deep optical images and sensitive optical spectroscopies with the future instruments may provide key measurements (e.g. the kinematics, the chemical abundances and the ages of stars) to better parametrize and quantify the role of different processes. Simulations predict that stars in the disc's outskirts will have a high velocity dispersion if formed with newly accreted gas \citep{Minchev12}, and will have slow circular rotation and a high radial velocity dispersion if migrated there via the scattering effects of bars \citep{Herpich17}. Stars migrated due to drivers other than bars will tend to have higher circular rotation and cooler dynamics \citep{Debattista17}. The stars formed with accreted gas should have different ages and chemical abundances from the migrated stars. In addition to that, we highlight the useful information from the H{\textsc I} data, and emphasize that due to the data depth our discussion is limited to the star-forming galaxies with detectable H{\textsc I} gas at the sensitivity of the ALFALFA data.We look forward to the sensitive and highly resolved H{\textsc I} images from the coming surveys like Apertif and WALLABY  \citep{Koribalski12}, which will significantly enlarge the size of the H{\textsc I} sample, allow for more direct comparisons between the optical and H{\textsc I} radial distributions, and provide more direct measurements of the spin parameters. 

 A dedicated component decomposition of galaxies (based on GALFIT \citep{Peng02} or other softwares) that simultaneously account for the bars, bulges and disc breaks will shed light on how the formation of the outskirts is coupled to that of the whole galaxy. A direct comparison with cosmological simulations of galaxy formation will be greatly helpful. The special population of H{\textsc I}-rich and low $\lambda$ galaxies identified in this paper provides a touchstone for the simulations, and such theoretical investigations are underway by the completion of this paper (Lagos et al. in prep, Fu et al. in prep).

\section*{Acknowledgements}
We thank the anonymous referee for constructive comments. 
We gratefully thank H. Gao, J. Fu, G. Kauffmann, P. Serra for useful discussions.

C. Lagos is funded by a Discovery Early Career Researcher Award (DE150100618) of the Australian Research Council.
C. Li acknowledges the financial support of the National Key Basic Research Program of China (No. 2015CB857004) and NSFC grants (No. 11173045, 11233005, 11325314, 11320101002). Parts of this research were conducted by the Australian Research Council Centre of Excellence for All Sky Astrophysics in 3 Dimensions (ASTRO 3D), through project number CE170100013.

GALEX (Galaxy Evolution Explorer) is a NASA Small Explorer,
launched in April 2003, developed in cooperation with the
Centre National d'Etudes Spatiales of France and the Korean Ministry
of Science and Technology.

We thank the many members of the ALFALFA team who
have contributed to the acquisition and processing of the ALFALFA
dataset over the last many years. RG and MPH are supported by NSF
grant AST-0607007 and by a grant from the Brinson Foundation.

Funding for the SDSS and SDSS-II has been provided by
the Alfred P. Sloan Foundation, the Participating Institutions, the
National Science Foundation, the U.S. Department of Energy,
the National Aeronautics and Space Administration, the Japanese
Monbukagakusho, the Max Planck Society, and the Higher Education
Funding Council for England. The SDSS Web Site is
http://www.sdss.org/.

\bibliographystyle{mnras}
\bibliography{diskbreak}

\appendix
\label{sec:appendix}
\section{Examples of products in the photometric pipeline}
\label{sec:appendix_phot}
This section supplement with examples for steps of the photometric pipeline described in Section~\ref{sec:SDSSdata}.

\begin{figure} 
\includegraphics[width=7cm]{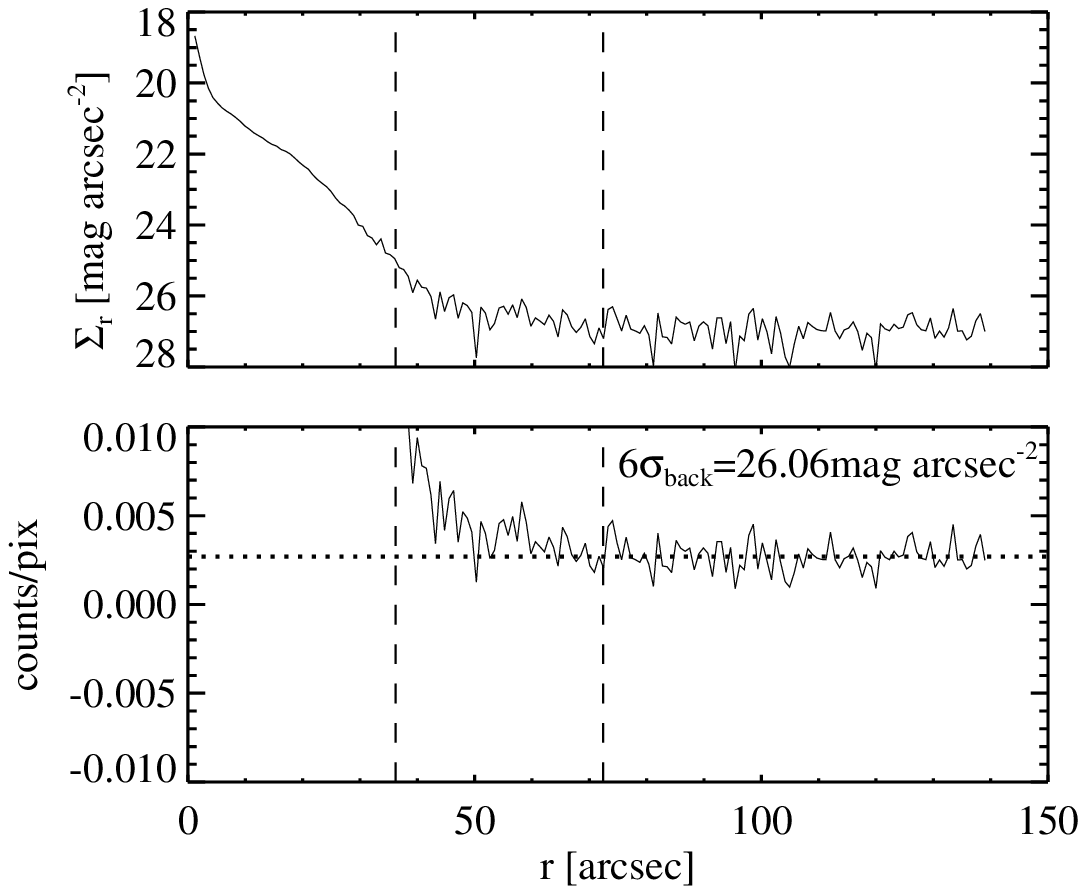}
\caption{An example of subtracting the residual background value from the SB profile. The two panels show the same $r$-band SB profile of an example galaxy before the subtraction of residual background (Section~\ref{sec:getSBprof}). The top panel shows the whole profile in the unit of mag arcsec$^{-2}$, and the bottom panel zooms in the outer faint part of the profile in the unit of counts per pixel. The dashed vertical lines mark the positions of $R_{25}$ and 2$R_{25}$. The outer part of the profile has reached the background of the image and is flat. We take the data points in the profile which have $r>2R_{25}$, and estimate the residual background as the 3$-\sigma$ clipped mean value of these data points. The dotted horizontal line mark the estimated residual background. The image depth is estimated as $6\sigma_{back}$.}
\label{fig:example_backsub}
\end{figure}

As described in Section~\ref{sec:getSBprof}, we estimate the background value of each image in two steps. We first mask all the detected sources and obtain a rough estimate of the background value from the image. Then we derive the SB profile of the target galaxy in the image out to the radius $4R_{25}$. The flattened outer part of the SB profile (beyond $2R_{25}$) is used to estimate the residual background value. Figure~\ref{fig:example_backsub} provides an example of the second step of subtracting the residual background value.  We also estimate the image depth (the faintest SB that can be reliable measured) as $6\sigma_{back}$ of data points in the flattened part of the profile. 

\begin{figure} 
\includegraphics[width=7cm]{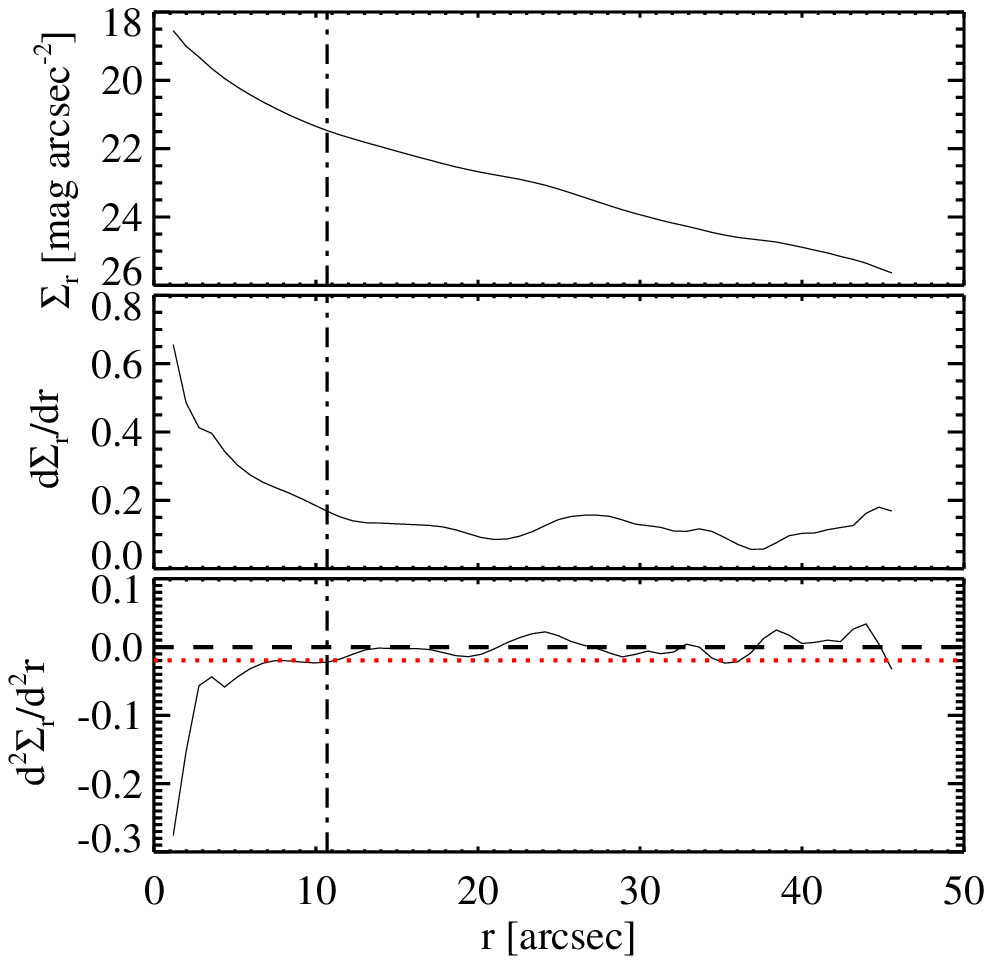}
\caption{An example of identifying the bulge dominated region from the SB profile. The top, middle and bottom panels show the smoothed SB, the first-order derivative ($\frac{{\rm d}\Sigma}{ {\rm d} r }$), and the second-order derivative ($\frac{{\rm d^2}\Sigma}{ {\rm d^2} r }$) profile of the same galaxy respectively. In the bottom panel, the dashed horizontal line marks zero and the red dotted horizontal line marks minus the standard deviation of $\frac{{\rm d^2}\Sigma}{ {\rm d^2} r }$  ($\sigma_{deriv}$). The dash-dotted vertical lines mark $r_{in}$, beyond which the curve of $\frac{{\rm d^2}\Sigma}{ {\rm d^2} r }$ rises above the red dotted line. They serve as the the division between bulge and disc dominated regions. } 
\label{fig:example_bulge}
\end{figure}

As described in Section~\ref{sec:bulgemask}, we use the second-order derivative profile to identify and mask central regions that are dominated by bulges. We give an example of this procedure in Figure~\ref{fig:example_bulge}.

\begin{figure*} 
\includegraphics[width=5.8cm]{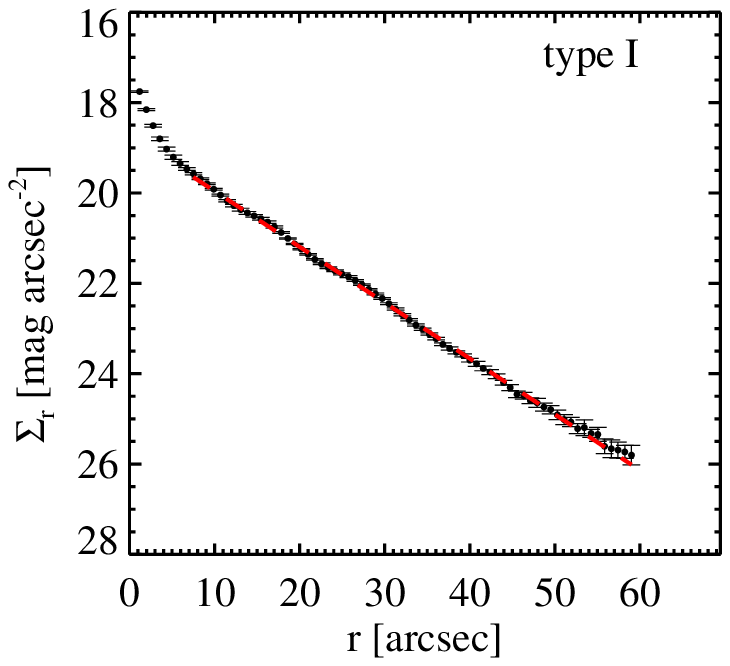}
\includegraphics[width=5.8cm]{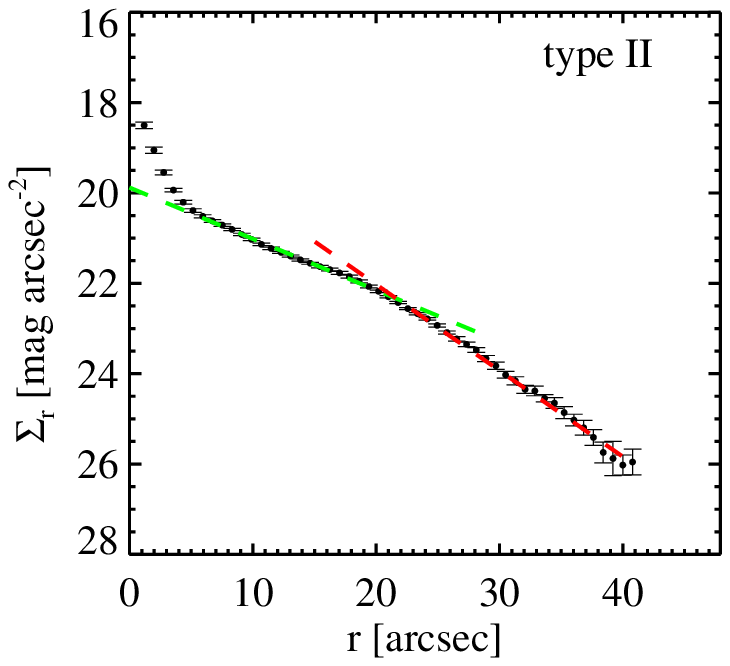}
\includegraphics[width=5.8cm]{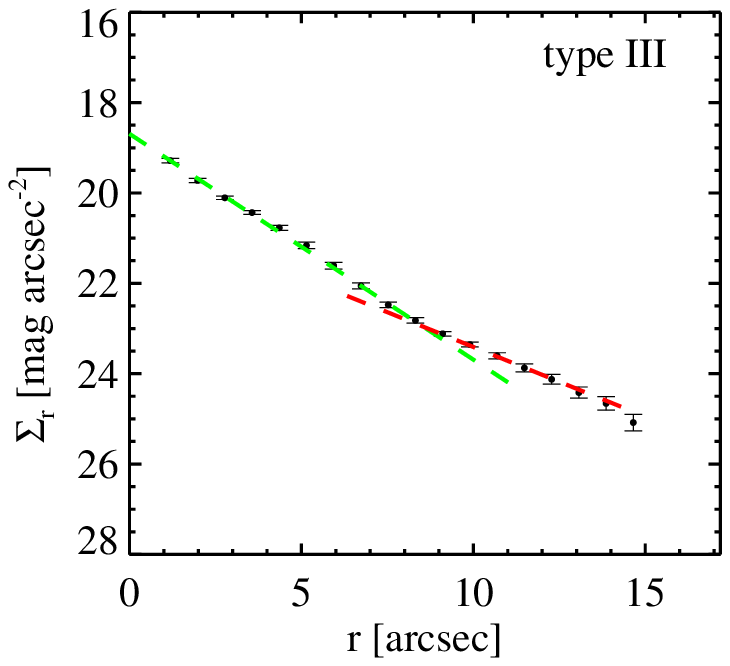}

\includegraphics[width=6cm]{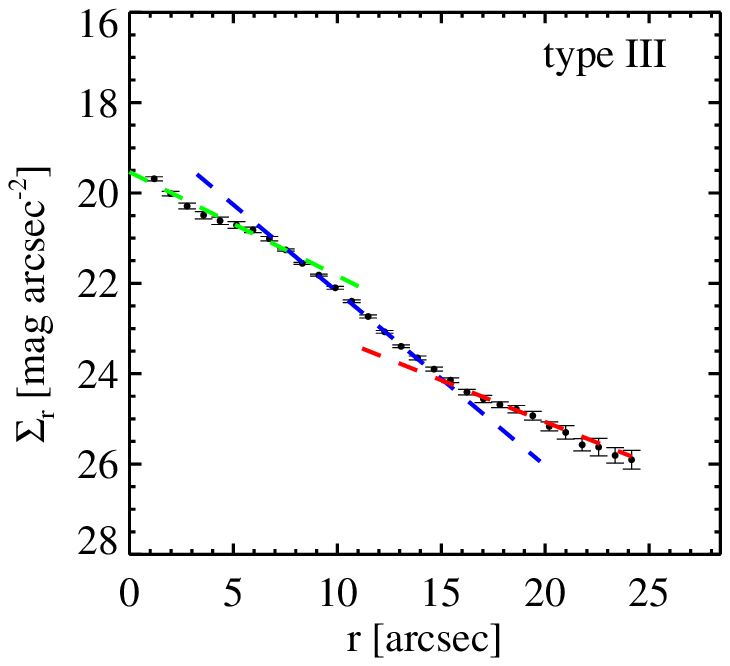}
\includegraphics[width=6cm]{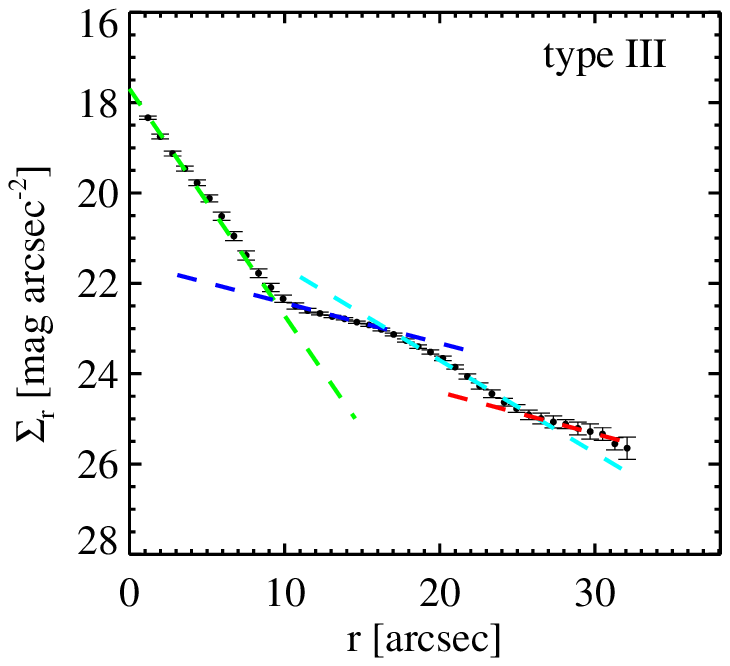}
\caption{Examples of individual SB profiles identified to host type I-III discs. The top row shows three galaxies which each has only one break in the disc. The bottom row show two type III galaxies with two and three breaks in the discs respectively. The observed SB profiles are displayed in dots and error bars. For the type I galaxy, the red line show the best-fit model. For the type II and III galaxies, the red lines show the outmost discs beyond the break radius from the best-fit models, and the lines in other colours show the broken inner discs. } 
\label{fig:example_prof}
\end{figure*}

We identify the outmost disc break from each SB profile in Section~\ref{sec:breakclassify}. In the top row of Figure~\ref{fig:example_prof}, we show examples of type I-III galaxies which each has only one disc break. We also show type III galaxies which each have more than one disc breaks (bottom row). As described in Section~\ref{sec:breakclassify}, for each SB profile, we iteratively search for breaks outward until reaches the radius which has a distance of 5.2 arcsec from $r_{out}$. We use the property of the outmost disc break to classify the galaxy. 

\section{Properties of the breaks}
\label{sec:appendix_breakprop}
\begin{figure*} 
\includegraphics[width=14cm]{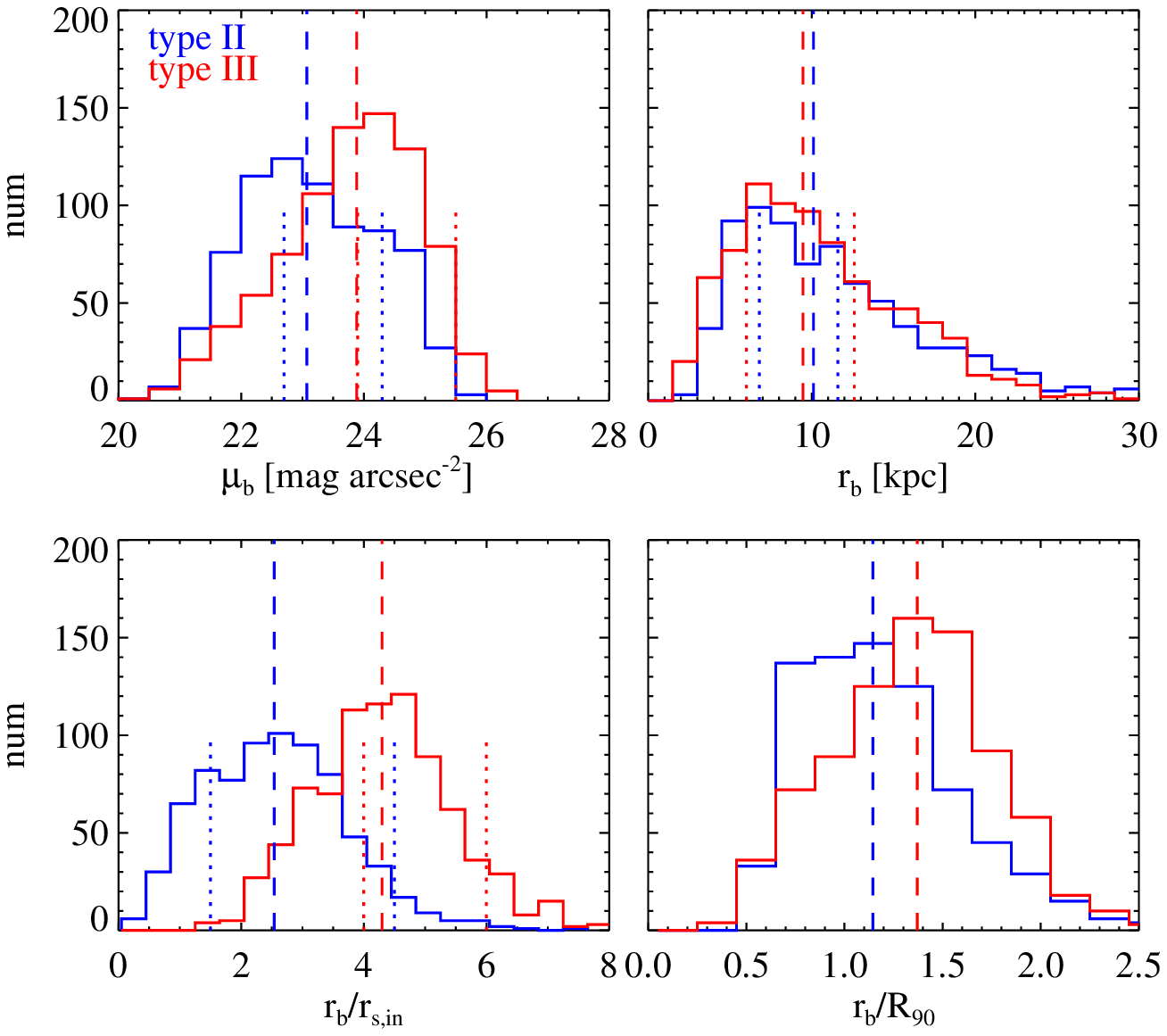}
\caption{Distributions of type II and III break properties. The dashed lines mark the median value of each distribution. The dotted lines mark the range of distributions from \citet{Pohlen06}.}
\label{fig:breakprop}
\end{figure*}

This appendix reproduces some well-known characteristic features of type II and III disc breaks.
Figure~\ref{fig:breakprop} shows the distributions of break related properties measured in this paper. The top and bottom-left panels compare the distributions of $\mu_b$, $r_b$ and $r_b/r_{s,in}$ with the typical range of distributions from \citet{Pohlen06}. Although the exact ranges are not the same, the general trends of the differences between type II and III galaxies from our data are consistent with \citet{Pohlen06}: type II galaxies on average have brighter $\mu_b$, slightly smaller $r_b$ and smaller $r_b/r_{s,in}$ than the type III galaxies. In the bottom-right panel, $r_b/R_{90}$ are typically close to unity for type II galaxies and larger than one for type III galaxies, consistent with previous studies (Z15).

\begin{figure*} 
\includegraphics[width=5cm]{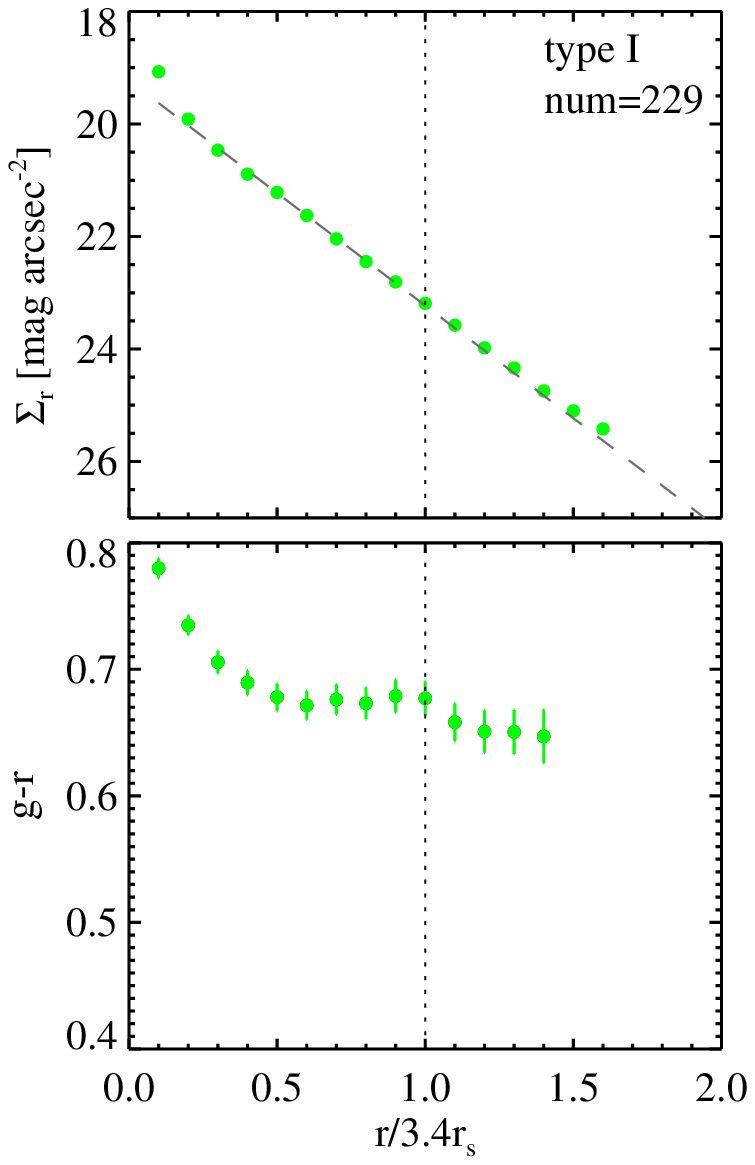}
\includegraphics[width=5cm]{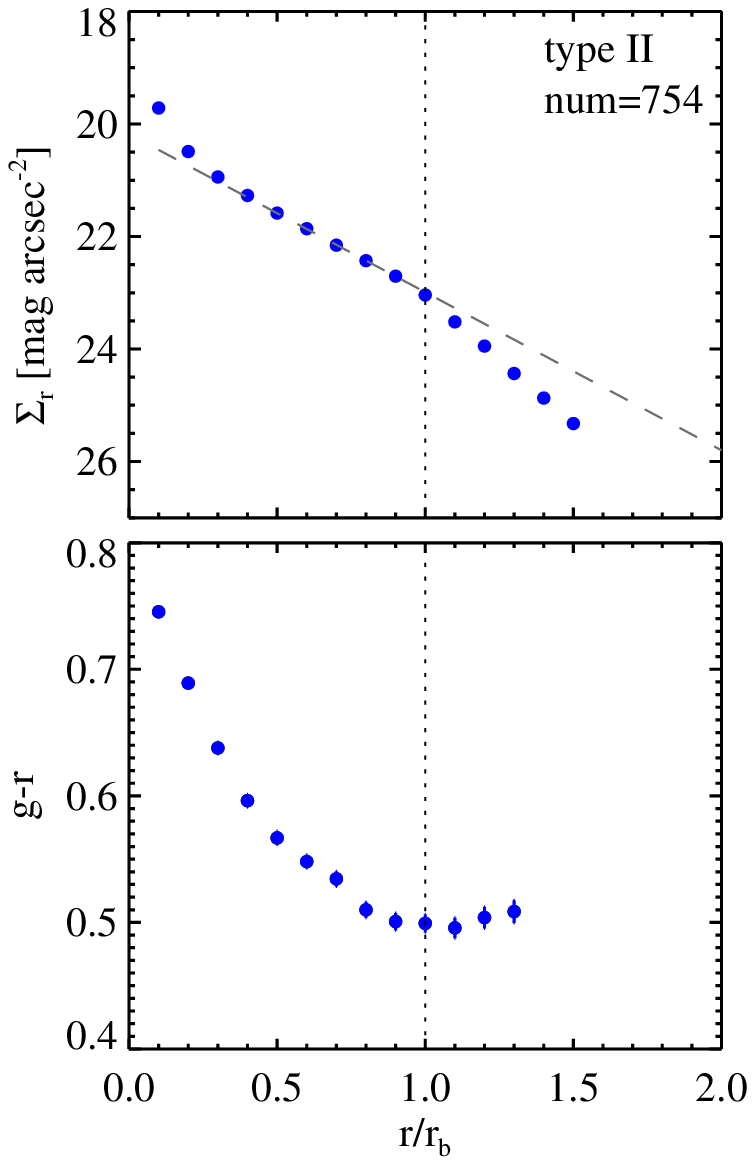}
\includegraphics[width=5cm]{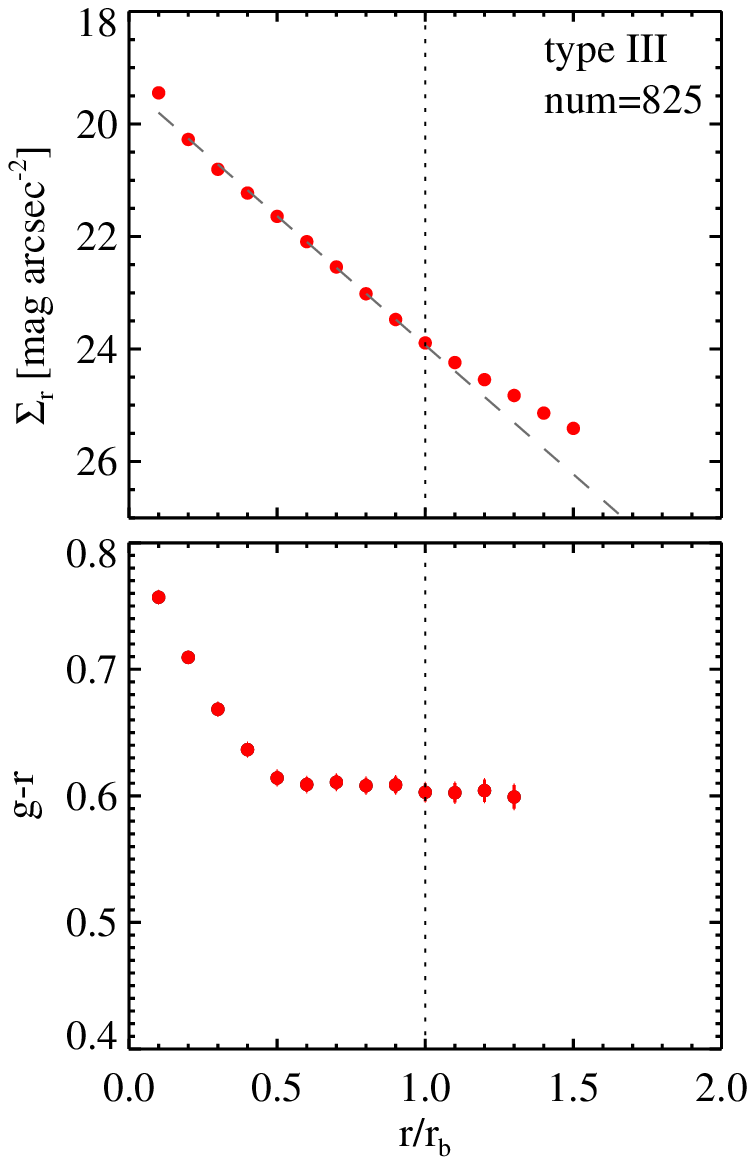}
\caption{The median surface brightness and colour profiles for type I-III galaxies. The surface brightness profiles are in the top row and the colour profiles are in the bottom row. The left, middle and right columns are for type I, II and III galaxies respectively.  The dashed lines are linear fits to the median profile in the radius range 0.5-1 $r_b$. The dotted lines mark $r_b$. The error bars show standard errors of the median values. The surface brightness profile of each individual galaxy stops where the error is larger than 0.18 mag (6$\sigma$), and the colour profile of each individual galaxy stops where the error is larger than 0.2 mag. Only when more than 60\% of the galaxies have reliable measurements, do we calculate the median value in each radius bin.}
\label{fig:colorprofile_all}
\end{figure*}

Figure~\ref{fig:colorprofile_all} shows the averaged surface brightness profile and $g-r$ colour profile for type I-III galaxies. The shape of the colour profiles, especially the U-shaped upturn near $r_b$ for type II galaxies, are consistent with previous results \citep{Bakos08}. 

\end{document}